\documentclass[fleqn,trackchanges,twocolumn]{aastex701}

\usepackage{amsmath}

\begin{document}

\title{Reconnection-driven State Transitions in Flat Spectrum Radio Quasars}

\author[orcid=0000-0003-1101-8436]{Agniva Roychowdhury}
\affiliation{National Centre for Radio Astrophysics - Tata Institute of Fundamental Research, Pune University Campus, Ganeshkhind, Pune 411007, MH, India}
\email[show]{agniva.physics@gmail.com}  

\begin{abstract}
We extend the work of Roychowdhury (2026) on skewness variations of the logarithmic flux, driven by large GeV flares in FSRQs, to a sample of 18 FSRQs. We find that they can be categorized into three groups, one where the skewness attains a persistent lower value after a large flare, one where it increases, and those where change in skewness is not significant. To provide a theoretical ground for these results, we use the statistical plasmoid model of Fermo et al. (2010) that self-consistently produces large plasmoids through merging which, when gain energy from the reconnection event and are Doppler aligned, produce large flares. We find that a downsampling of our simulation of 1500 runs to 18 statistically reproduces the observed distribution in p-values for change in skewness. We further compute the ensemble Shannon entropy of the system and the skewness, where the entropy is found to decrease at a $3\sigma$ level in both the groups where skewness either increases or decreases, as a direct evidence of increase in order in the system caused by a flare. We find that the power spectral densities of the simulated light curves are broken-power-laws, resembling a white noise+red noise broken by the typical cooling timescale in our system, in accordance with known blazar variability. We find that our results are robust to a $200-300\%$ change in several fiducial parameters of the simulation. Our stochastic simulation of plasmoids inside a blazar jet is consistent with key observable statistical properties of blazar GeV light curves.

\end{abstract}

\keywords{galaxies: active - galaxies: jets}


\section{Introduction} 

Blazars, a subclass of active galactic nuclei, show extreme variability (minutes to years) in multiple wave-bands, especially in the $\gtrsim$ GeV range \citep[e.g.]{gian09,raiteri25}. The established statistical trends in blazar variability are commonly found in time series analyses in numerous fields of science, mostly accentuated by pink/red noise-like power spectral densities found in electronics, naturally occurring systems and even the brain \citep[e.g.]{schottky25,btw87,halley96, voytek15}. Since the power spectral density only captures the \textit{overall} behaviour of the variability, there is a significant amount of degeneracy as far as its actual origin is concerned. In blazars (optical-GeV) the fundamental mechanism to produce red noise is either continuous turbulent cascades where stochastic fluctuations in the local magnetic field and electron density are integrated over a multi-zone emission region \citep{marscher14} or a series of shot noise profiles which can be produced by a series of shock waves travelling down the jet \citep{marsch85, mukherjee19} randomly initiated in time. However, in calculations of PSDs from multi-wavelength light curves separated in time, it is often assumed that the blazar produces variability that exhibits stationarity, i.e., the mean and variance are constant in time.

Furthermore, it was previously established that blazar light curves, especially in X-rays and higher energies, exhibit a log-normal-like flux distribution, that arises from multiplicative processes within the jet. It turns out such distributions can arise through appropriate time-dependent modulations in the electron kinetic equation \citep{sinha18} or situations where the logarithms of the multiplicative factors to the flux follow a normal distribution. However, a self-consistent prescription was missing. Very recently \cite{royc25}, using long-term 18 year Fermi-LAT light curves of a famous flat spectrum radio quasar (FSRQ) CTA 102, deduced that not only does the flux distribution deviate appreciably from a log-normal distribution throughout the entire interval, but the largest flare in 2017 ``resets" the system and relaxes the distribution to a more steady-quiescent state with a statistically significant lower skewness. They explained it with a modified minijets-in-a-jets formalism of \cite{biteau12} and found that the largest GeV flares are rare events occurring due to Doppler aligning of a maximum number of minijets, which can be followed by a state change of the blazar. If the minijets are emitting plasmoids heated by reconnecting events, the state change could be qualitatively explained by magnetic relaxation, where the magnetic energy reduction due to the large flare reduced the number of reconnecting events, and hence energy dissipation was reduced, resulting in a stabler internal configuration of the bulk jet.

This is partly in accordance with many papers on reconnection-driven high-energy variability in quasar jets \citep{giannios13,petropolou16,petrop18,sironi25,das26}, where tearing instability in a highly magnetized current sheet spontaneously creates a series of plasmoids, which when merge together release energy into and form a ``monster" plasmoid, which is responsible for large GeV flares \citep{giannios13}.

In this work we add to CTA 102 seventeen more Flat Spectrum Radio Quasars (FSRQs) from the Fermi-LAT light curve repository as a pilot-study to extend the statistical analyses of \cite{royc25} to investigate if flare-induced state transitions could indeed be a physical property of blazars or the case of CTA 102 was only anomalous. To better the understand the observations, we use a self-consistent statistical model of plasmoid mergers from \cite{fermo2010} that uses a Smoluchowski-like coagulation equation to track the time evolution of an initial distribution of magnetic islands through injection, escape and merging. In essence, we provide a general theoretical framework for high-energy blazar variability and compare simulation results to FSRQ data reduced in this work. 

Section \ref{sec:obs} discusses the FSRQ data and related analyses. Section \ref{sec:sim} shows the simulation results and maps it to observational results. Sections \ref{sec:disc} and \ref{sec:conc} discusses the global implications of our study.

\section{Observations and Data Analyses}
\label{sec:obs}
\subsection{GeV Light Curve Analysis of Eighteen FSRQs including CTA 102}

We used the Fermi-LAT light curve repository to download full GeV light curves of a total of eighteen FSRQ-type blazars, hand-picked for a pilot study and which do not represent a statistically complete sample of any sort. They are listed in Table \ref{tab:sample} with corresponding popular names. All of them are highly flaring blazars with 18 year GeV variability data. We downloaded weekly variability data and made a cut of TS $\geq 25$ on each of the light curves for our analysis.

\begin{table}[ht!]
\centering
\caption{Expanded sample of 18 flaring FSRQs}
\label{tab:sample}
\begin{tabular}{lll}
\hline \hline
 & 4FGL Identifier & Popular Name \\ \hline
1  & J0108.6+0134 & 4C +01.02 \\
2  & J0210.7$-$5101 & PKS 0208$-$512 \\
3  & J0221.1+3556 & B2 0218+35 \\
4  & J0339.5$-$0146 & CTA 026 \\
5  & J0403.9$-$3605 & PKS 0402$-$362 \\
6  & J0428.6$-$3756 & PKS 0426$-$380 \\
7  & J0457.0$-$2324 & PKS 0454$-$234 \\
8  & J0538.8$-$4405 & PKS 0537$-$441 \\
9  & J0730.3$-$1141 & PKS 0727$-$11 \\
10 & J1159.5+2914 & Ton 599 \\
11 & J1224.9+2122 & 4C +21.35 \\
12 & J1229.0+0202 & 3C 273 \\
13 & J1256.1$-$0547 & 3C 279 \\
14 & J1427.9$-$4206 & PKS 1424$-$41 \\
15 & J1512.8$-$0906 & PKS 1510$-$089 \\
16 & J1833.6$-$2103 & PKS 1830$-$211 \\
17 & J2232.6+1143 & CTA 102 \\
18 & J2253.9+1609 & 3C 454.3 \\ \hline
\end{tabular}
\end{table}

\subsubsection{The Rolling Skewness Diagnostic}

\begin{figure*}
    \centering
    \includegraphics[width=0.8\linewidth]{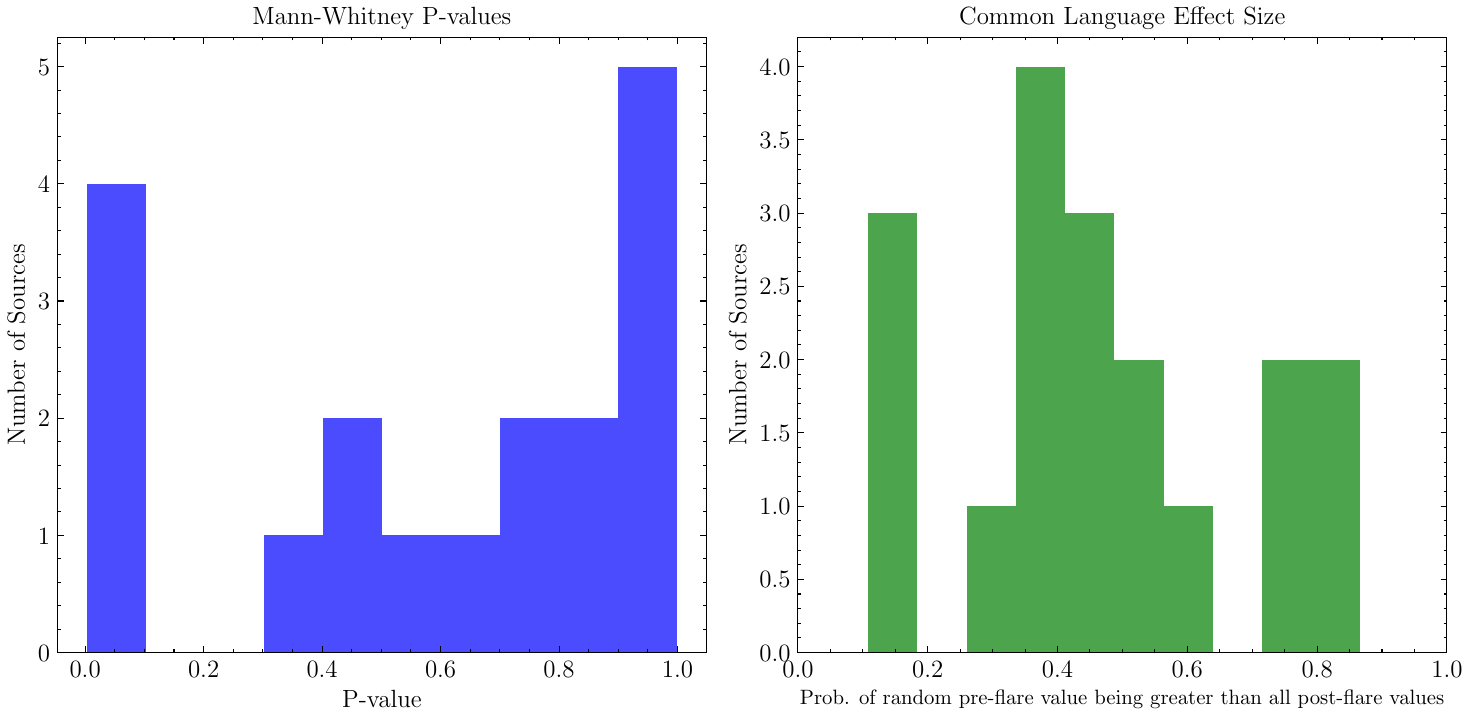}
    \caption{Left panel : Figure shows the histogram of p-values obtained from the Mann Whitney U Test between the preflare and postflare skewness distributions of the eighteen FSRQs in our sample. The departure from unimodality is not statistically significant. Right panel : The corresponding common language effect size using the Mann Whitney U statistic. More clutter at $<0.5$ is evident, in expectation with the p-value histogram.}
    \label{fig:pvalue_skew}
\end{figure*}

We first identify the position of the \textit{largest} flare in the system. This does not take into account if there are similar magnitude flares before and after that flare, and in the process introduces simplicity until further detailed analysis is required. After identifying the flare peak, following the method of \cite{royc25}, we take a rolling skewness window of size $\sim1$ year, and slide it from a maximum of $\sim 6$ years \textit{before} the flare to $\sim6$ years \textit{after} the flare. It is likely that there are not enough points in the light curve to sample that time window and hence the skewness values do not exist in those times. However, it is not possible to ensure a fully fair comparison since blazar flaring is a priori unpredictable. To this end, we chose the time window such that at least 50\% of our sources have data at the edges of the window. The pre-flare and post-flare segments are defined keeping a 2 week buffer around the flare, so as to not include the natural mathematical spike in skewness due to a flare. The rolling skewness values are inherently not \textit{independent}, since at each movement across the light curve, one point is added and one removed, while all remaining points are re-used to compute the skewness. We determined the corresponding auto-correlation function $\rho(t)$ of the rolling skewness, both in the pre-flare and post-flare segments, and used the integrated autocorrelation time to estimate the \textit{effective} number of \textit{independent} samples in the rolling skewness. This is given by $N_{\rm indep}=N/[1+2\Sigma_i \rho(t_i)\Theta(\rho(t_i))]$ where $N$ is the number of rolling skewness points, the denominator is the integrated autocorrelation time $\tau$ and $\Theta$ is the Heaviside function. Then $N_{\rm indep}$ values of the corresponding rolling skewness sets (in pre and post flare) were chosen out of the total rolling skewness in steps of $\tau$. For cases where there was no data in the time-window, the skewness value at the edge of the observed window was extended to the analysis window. For each source, we then computed the Mann Whitney U Test \citep{mw47} between the pre-flare and the post-flare skewness (up to $60$ days before and after the flare) to confirm if the skewness has decreased at a statistically significant level. We further computed the corresponding distribution of the common language effect size ($C=U/n_1n_2$, where $U,n_1,n_2$ are the Mann Whitney U statistic and the sizes of the pre and post flare rolling skewness samples after auto-correlation correction) for each source, that serves as an alternate representation of the same statistical method. It directly tells us the \textit{fraction} of cases where a randomly chosen skewness value from one sample is larger than all the skewness values in the other sample, which in this case is the pre-flare and post-flare skewness sets respectively. Figure \ref{fig:pvalue_skew} shows the resulting p-value distribution of the skewness reduction for all the sources.

The p-value distribution is spread through the entire range, with more concentration around $>0.9$. A simple Hartigan's Dip Test \citep{hartigan1985} provided a statistic $\sim0.1$ at p-value $0.1$, implying a marginal evidence of unimodality violation. The common language effect size histogram is similarly scattered and is more cluttered around $\lesssim0.5$, providing a proxy for the p-value histogram, with majority of sources at the higher p-values, implying that the probability of skewness being higher in the pre-flare states than the post-flare is low in these sources.

It is worthwhile to examine the edge cases of the p-value histogram. For the lowest p-values, these indicate sources where the skewness has reduced after the largest flare, implying the source has reached a steady/quiescent state with less dominant \textit{outlier} events in the luminosity distribution than the preflare. In other words, the large flare has \textit{relaxed} the system reducing the total magnitudes of flares, or the departure from a symmetric luminosity/flux distribution. Second, where the skewness has increased (largest p-values), the largest flare has either \textit{increased} the flaring making the system more unstable or it was not a \textit{special} flare, implying the magnitudes of outlier events (compared to the baseline luminosity post-flare) have increased, regardless of the flaring rate. The remaining sources with moderate p-values are unclassifiable hence, where it is unclear if the largest flare had \textit{any} role to play in physically changing the underlying flare production rate in the blazar.

\begin{figure}
    \centering
    \includegraphics[width=\linewidth]{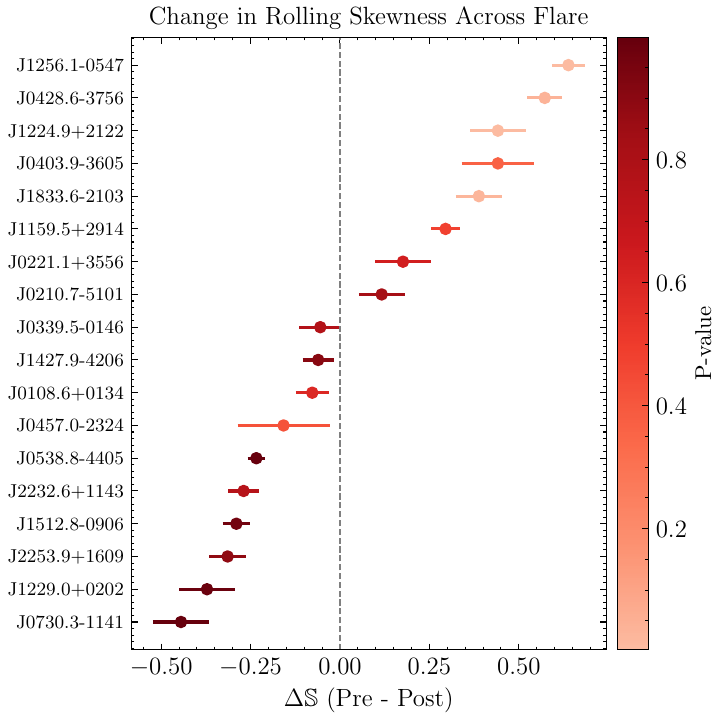}
    \caption{Figure shows the change in \textit{median} skewness $\Delta\mathbb{S}$ through the largest flare for each of our sources, colored by the corresponding p-values for the statistical significance of a skewness reduction across the flare. The distribution of $\Delta \mathbb{S}$ is smooth, between $\sim -0.5-0.75$, exhibiting the range of different physical properties of our sample.}
    \label{fig:del_skew}
\end{figure}

\begin{figure*}
    \centering
    \includegraphics[width=0.95\linewidth]{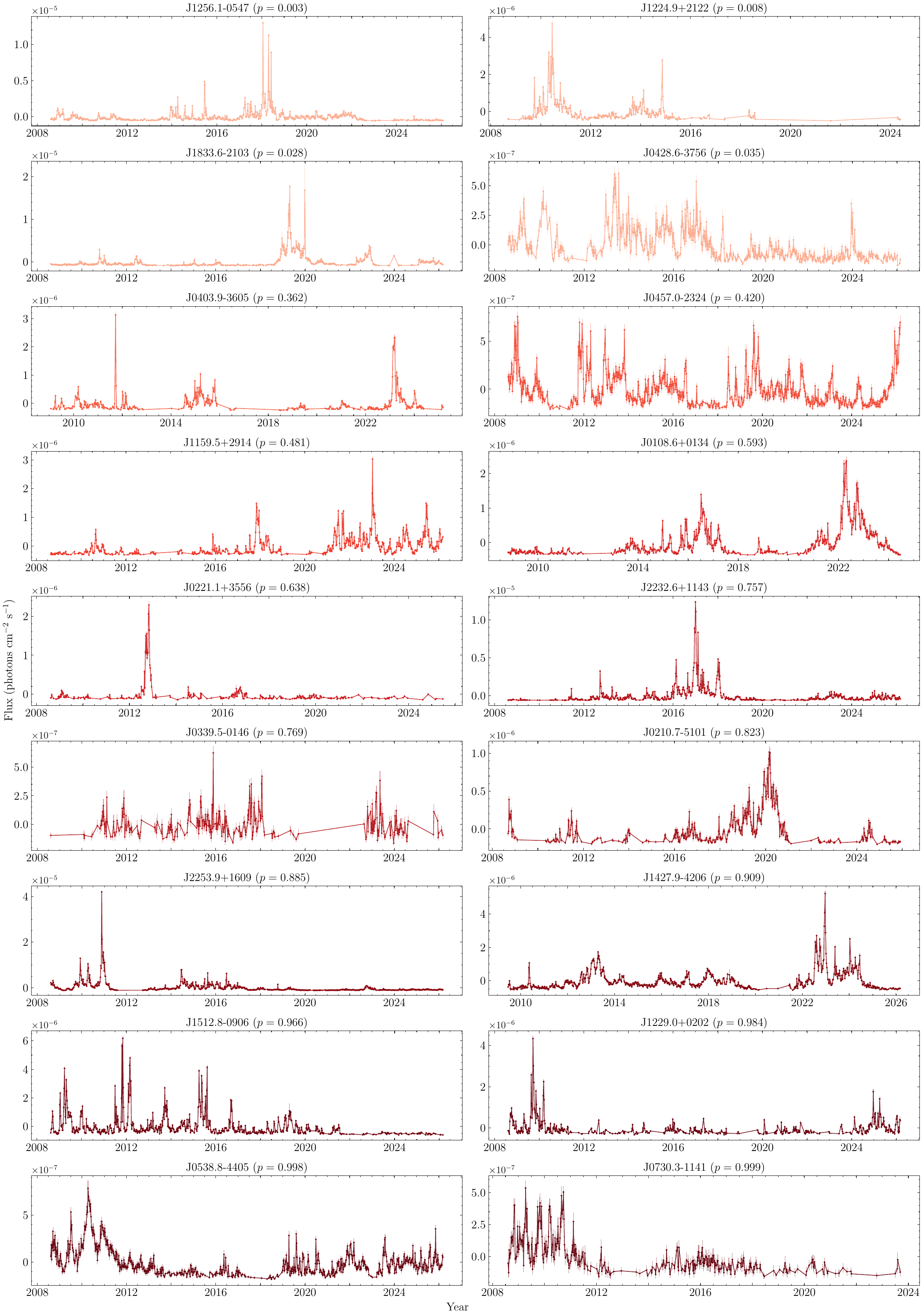}
    \caption{Light curves arranged in order of increasing p-value, with lightest light curves showing a statistically significant reduction in skewness and boldest light curves showing a statistically significant increase in skewness.}
    \label{fig:sig_nonsig}
\end{figure*}

Figure \ref{fig:del_skew} shows the corresponding change in \textit{median} skewness $\Delta \mathbb{S}$ between the pre and post-flare segments for each source, colored by the corresponding p-value for the statistical significance of the skewness decrease across the flare. A smooth transition between the edges of the p-value histogram is discerned, with the sample exhibiting a full range of possibilities, one where skewness decreases, one where skewness change is not conclusive and the other where the skewness had increased, after the largest flare. Using the same, Figure \ref{fig:sig_nonsig} shows the light curves in increasing order of p-values, where the lightest and boldest show a statistically significant decrease and increase in skewness across the flare respectively. We note here that CTA 102, i.e., J2232+1143, shows a moderate p-value, and it borders on an increase in skewness, contradicting what was described in \cite{royc25}. This is because \cite{royc25} did not use the auto-correlation correction to correct for dependent samples and hence the statistical significance of CTA 102's skewness reduction was overestimated.

\section{Simulation and Results}
\label{sec:sim}
In this section, we will introduce a statistical model of plasmoids from the literature that incorporates the injection of new plasmoids into the system, their escape and their merging. Merging is relevant since that is the only way a \textit{monster} plasmoid may be created through merging, or \textit{coagulation}, of smaller plasmoids. 

\subsection{Smoluchowski Coagulation of Relativistic Plasmoids : Fermo et al. 2010 Model Description}

While monster plasmoids have been identified as a potential source for large GeV flares in blazars, there is currently no self-consistent treatment that produces observationally testable variability in the system. To this end, we use in this work a robust one-dimensional statistical model of plasmoid evolution in current sheets derived in \cite{fermo2010}. The model of \cite{fermo2010} derives a Smoluchowski-like coagulation equation \citep[e.g.]{smol16,davies99} to describe the evolution of the magnetic flux distribution of a set of plasmoids, as they merge to form bigger plasmoids and release energy into the merged plasmoid as their field lines reconnect. The equation, directly from \cite{fermo2010} reads :

\begin{equation}
\begin{aligned}
&\frac{\partial f}{\partial t} + \frac{\partial}{\partial \psi}(\dot{\psi} f) + \frac{\partial}{\partial A}(\dot{A} f) = S(\psi, A) - \frac{c_{A}}{L} f + \frac{1}{L} \\
&\int_{0}^{A} dA' f(\psi, A') \int_{0}^{\psi} d\psi' v(\psi, A', \psi', A - A') f(\psi', A - A') \\
&- \frac{1}{L} f(\psi, A) \int_{0}^{\infty} dA' \int_{0}^{\infty} d\psi' v(\psi, A, \psi', A') f(\psi', A') 
\end{aligned}
\label{eqn:fermo}
\end{equation}

where $f(\psi, A, t)$ is the distribution function of plasmoids per unit flux per unit area at time $t$, and $\psi$ and $A$ denote the magnetic flux and cross-sectional area of an individual plasmoid, respectively. The flux and size of an individual plasmoid increases in time with the relation $\dot{\psi}=\varepsilon c_AB_0$ and $\dot{r}=\varepsilon c_A$ respectively, due to \textit{global} magnetic reconnection, where $B_0$ is upstream magnetic field fed continually into the current layer. $c_A$ is the characteristic Alfven speed that defines the speeds of these plasmoids and $B_0$ is the magnetic field strength in the sheet. The left hand side of the equation governs the growth of the flux distribution function as a function of its dependent variables. The right hand side contains a source term (injection of new plasmoids into the system) $S(\psi,A)$ where $A=\pi r^2$, a sink term $-c_A f/L$ describing the advection of plasmoids out of the system, and two terms where the first describes the creation of a new plasmoid formed due to merging of two plasmoids and the second describes a loss of a plasmoid due to merger with another. $v$ is the relative velocity of two plasmoids, which essentially controls the probability of merging. $v$ can be written as :

\begin{equation}
v^2(\psi_1, A_1, \psi_2, A_2) = \frac{\varepsilon^2}{4\pi\rho} \frac{\psi_1 \psi_2 r_1 r_2 (r_1^2 + d_i^2)^{1/2} (r_2^2 + d_i^2)^{1/2}}{(r_1^2 + d_e^2)^{3/2} (r_2^2 + d_e^2)^{3/2}}
\end{equation}

Equation \ref{eqn:fermo} is a non-linear integro-differential equation, making it analytically intractable and even numerically difficult to solve without simplifications. Further, Equation \ref{eqn:fermo} is deterministic, that must have a clean numerical solution at every point in time. It turns out that this class of Smoluchowski coagulation equations can be solved much more directly using Gillespie's Algorithm, which is essentially a Monte Carlo method to solve such equations \citep[e.g.]{gillespie76,gillespie77,anderson15}. Since the total dynamics is governed by the injection, escape and merger terms on the right hand side of Equation \ref{eqn:fermo}, one can use a total Gillespie rate parameter $\lambda$, which can be summed as $\lambda=\lambda_{\rm inj}+\lambda_{\rm esc}+\lambda_{\rm merge}$, where the first two terms will simply be $(\psi,A)$ and $c_A N/L$ respectively, where $N$ is the number of plasmoids present in the region at any given time. The merging terms, in the most basic algorithm, need to be handled on a brute-force pair-wise basis. However, one can bypass such a method using a thinning algorithm from \cite{lewis79} (also see \citealt{ogata81,brown2005}). One first derives the maximum merger rate of the system by assuming \textit{all} pairs will be merging. This can be written as $\lambda_{\rm merge}=N(N-1)v_{\rm max}/2L$, where $v_{\rm max}$ is given by

\begin{equation}
v_{max} = \sqrt{\frac{\varepsilon^2}{4\pi\rho}} \frac{\psi_{max} \cdot r_{max} \cdot \sqrt{r_{max}^2 + d_i^2}}{(r_{max}^2 + d_e^2)^{1.5}}
\end{equation}

However, since this is an overestimation of the merger rate, one must \textit{delete} those instances where the random number $X$, drawn from $\mathcal{U}(0,1]$, satisfies $X\geq v_{\rm ij}/v_{\rm max}$ by accepting those only with $X<v_{\rm ij}/v_{\rm max}$. This essentially captures the same brute force rate, but at a much reduced computational complexity.

The bigger importance of the Gillespie algorithm stems from its ability to capture the inherent stochastic nature of systems like this, and hence in many cases preferred to the deterministic solution. The full procedure can be summarized as follows. At every time step, one draws a random number from $X=\mathcal{U}(0,1]$ and calculates the total rate $\lambda$. Depending on the relative magnitudes of the three rates (and hence probabilities of the three events), the system will proceed to inject a plasmoid or magnetic island into the current layer, advect a random island out of the system or merge any two islands randomly. Note that none of them can happen simultaneously in this algorithm. Further, the future time step $\Delta t$ is determined by the total rate $\lambda$ as $\Delta t=-\ln X/\lambda$ at the previous time step \citep{gillespie76}.

The above captures the entire process of  the plasmoid flux distribution evolution. From \cite{fermo2010}, when two plasmoids merge, the flux of the larger plasmoid is retained, while the flux of the other is dissipated into the plasma through reconnection. To incorporate this, we have allowed for this energy to heat up emitting particles in the newly formed plasmoid, where a large plasmoid merger will naturally produce more energy than a smaller merger. Part of this energy ($\Delta E\propto \psi^2$) is converted into luminosity, with a phenomenological synchrotron-like cooling factor $\exp[-\kappa_{\rm cool}B_0^2t]$, that would naturally produce an instantaneous luminosity rise with a decaying cooling profile. To incorporate this self-consistently into the entire system, we add a fraction $f_{\rm gain}\Delta E$ to the internal energy, or equivalently the luminosity of the plasmoid. It is expected that $f_{\rm gain}<1$ as all of the reconnection's dissipated energy will heat electrons and ions both.

We add the assumption that these plasmoids should not have a single orientation in the jet frame, or rather be isotropically directed. Upon merging, we employ the simplest prescription that the orientation of the larger plasmoid is retained since it must have a larger mass density and magnetic flux. Further, it is not a priori clear what the resulting Lorentz factors $\Gamma_{\rm mj}$ should be of each plasmoid. To first order, we fixed the Lorentz factors of all initial plasmoids, and did not evolve it due to mergers. This was done to keep the number of free parameters to a minimum. Given this scenario, the total luminosity of the plasmoids will be boosted by $\delta_{\rm mj}^3$ in the frame of the jet and the timescales will reduce/increase by $\delta_{\rm mj}$. To get the total observed jet luminosity, the total emission in the jet frame is multiplied by $\delta_{\rm jet}^2$ to maintain simplicity and prevent a complex electron heating prescription that would unnecessarily introduce many more parameters.

\subsection{Results}

\begin{figure*}
    \centering
    \hbox{\includegraphics[width=0.5\linewidth]{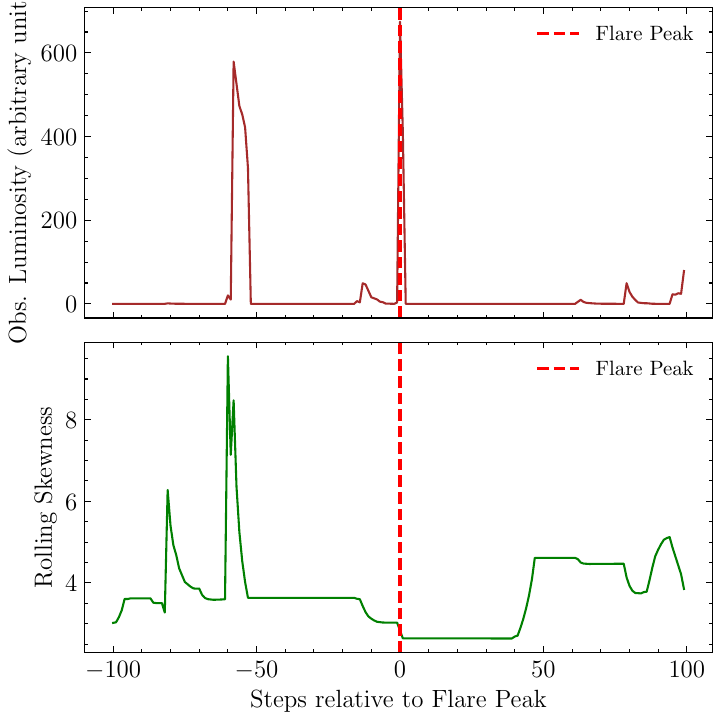}
    \includegraphics[width=0.5\linewidth]{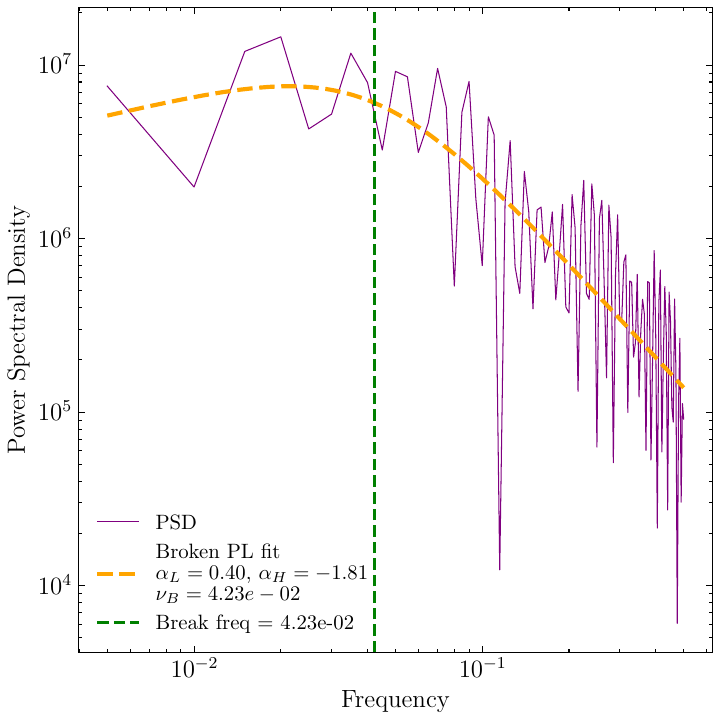}
    }
    \caption{Figure shows a simulated light curve from our model, that is consistent with real GeV flaring blazar light curves. The rolling skewness is shown in the bottom plot of the left panel, where it is discernible by eye that the skewness has decreased after the flare and stayed at that level for some time before again peaking due to new flares/plasmoid injection. The right panel shows the corresponding power spectral density of the light curve, that is a broken power law with a break frequency $\sim 0.04$ cycles/s and a pre and postbreak slope $\sim0.4$ and $-1.8$ respectively.}
    \label{fig:sync_lc}
\end{figure*}

\begin{table*}
\centering
\caption{Parameters of the Plasmoid Merger Model}
\label{tab:pars}
\begin{tabular}{l l l}
\hline
Parameter & Value & Description \\ \hline
\multicolumn{3}{l}{\textit{Global System Parameters}} \\
$L$ & 1000.0 & Characteristic length of the current layer \\
$B_{\rm 0}$ & 1.0 & Background magnetic field strength \\
$S_{\rm crit}$ & $10^4$ & Critical Lundquist number for injection triggering \\
$S_{\rm base}$ & 0.05 & Base injection rate for new plasmoids \\ \hline
\multicolumn{3}{l}{\textit{Reconnection \& Kinetic Parameters}} \\
$\varepsilon$ & 0.1 & Reconnection efficiency (rate of flux growth) \\
$c_A$ & 1.0 & Alfvén velocity (outflow speed) scaled to units of $c$ \\
$\kappa_{cool}$ & 0.05 & Cooling coefficient for internal energy decay \\ \hline
\multicolumn{3}{l}{\textit{Relativistic \& Energy Parameters}} \\
$\Gamma_{\rm mj}$ & $10.0$ & Lorentz factor of the plasmoids \\
$\Gamma_{\rm jet}$ & $20.0$ & Bulk Lorentz factor of the jet, with viewing angle $10^\circ$ \\
$\theta_{\rm mj}$ & $\mathcal{U}[-\pi,\pi]$ & Orientations of the plasmoids in the jet-frame \\
$\Theta$ & 1\% of the peak observed luminosity & Baseline luminosity from the bulk of the jet \\  
$f_{\rm gain}$ & 0.1 & Fraction of dissipated energy converted to internal energy \\ \hline
\multicolumn{3}{l}{\textit{Micro-physics \& Seed Scales}} \\
$d_i, d_e$ & 1.0, 0.1 & Ion and electron inertial skin depths \\
$\psi_{\rm seed}$ & 0.01 & Initial magnetic flux of injected plasmoids \\
$r_{\rm seed}$ & 0.1 & Initial radius of injected plasmoids ($0.1 \, d_i$) \\
$U_{\rm seed}$ & $10^{-4}$ & Initial internal energy per seed plasmoid ($\psi_{s\rm eed}^2$) \\ 
$N_0$ & 50 & Initial number of plasmoids \\ \hline
\end{tabular}
\end{table*}

In this section, we will discuss the results of our simulation. Table \ref{tab:pars} discusses all the model parameters for our simulation and the corresponding initial conditions. The simulation is initiated with $N_0=50$ plasmoids in a current layer of length $L=1000d_i$ and is allowed to evolve for a total of $5000$ time steps.  $c_A$ is taken to be close to speed of light (unity) for relativistic reconnection. Other parameters are generally free as long as certain constraints like the ratio of ion to electronic skin depth are not modified. The injection rate $S_{\rm base}$ determines $\lambda_{\rm inj}=S_{\rm base}B_0$ (simple magnetic field scaling added manually to make injections more rapid when $B_0$ is higher). One must note nevertheless that a large injection parameter ($S_{\rm base}$) will continuously inject plasmoids into the layer, making a very noisy light curve with many large flares. The situation is similar for the initial number of plasmoids too, and hence $N_0=50$ was chosen, that is in accordance with numerical simulations \citep{huang2010}. In contrast, a too low injection rate will reduce the probability of flares. We chose a $S_{\rm base}$ and $f_{\rm gain}$ that best captures the observed nature of blazar GeV variability. It is expected that $f_{\rm gain}$ must not be too high, since a huge part of the energy is dumped into ions/protons. Blazar jets are non-ideal and these two parameters capture that to a first order. In that respect $S_{\rm base}$ and $f_{\rm gain}$ were the only two \textit{free} parameters of our model that we varied. A further detailed discussion of the simulation parameters is not necessary for the scope of this paper since we are more interested in reproducing the observed trends in blazar GeV light curves given a set of physically possible parameters than in perfecting the choice of parameters for a purely simulation-based study. We refer the reader to Appendix \ref{app:A} for a sensitivity analysis of our results to a wide range of input parameters, where we find that the major results are relatively unchanged.

The total observed luminosity (in arbitrary time units) is given by:

\begin{equation}
\begin{aligned}
    L_{\rm obs}(t_{\rm obs})=\bigg[\sum_i U_{\rm i, mj}(t_{\rm mj}/\delta_{\rm mj})e^{-\kappa_{\rm cool}B_0^2t_{\rm mj}/\delta_{\rm mj}}\delta_{\rm mj}^{3}\bigg] \\
    (t_{\rm jet}/\delta_{\rm jet})\delta^2_{\rm jet}+1\% L_{\rm peak}
\end{aligned}
\end{equation}

where the first and last term on the right represent the total emission from the plasmoids (sum over internal energies $U$) and a baseline flux level from the jet bulk respectively. $t_{\rm mj}$ and $t_{\rm jet}$ represent the timescales in the frame of the plasmoid and the bulk jet respectively. Note that $(t_{\rm jet}/\delta_{\rm jet})$ is not a multiplicative factor, but that the total plasmoid emission in the jet frame is transformed to time units of the observation $t_{\rm obs}=(t_{\rm jet}/\delta_{\rm jet})$. A baseline flux is added in post-processing that is 1\% of the peak flux in a given run. We have explored the effect of increasing the same by at most 200 times in Appendix \ref{app:A} along with other parameters. We note that too high a baseline flux would submerge the stochastic properties of the light curve itself and hence a low-enough value that preserves a baseline and does not affect the statistical behaviour of the light curves was chosen.

Figure \ref{fig:sync_lc} shows a synthetic light curve from our simulation, the evolution of the rolling skewness (left panel), and the power spectral density for the same (right panel). The left panel of Figure \ref{fig:sync_lc} consistent with the general behaviour of GeV flaring blazars as in Figure \ref{fig:sig_nonsig}. The red dotted line shows the position of the flare and the rolling skewness is plotted in the same figure. The right panel shows the power spectral density of this light curve, showing a clear broken power law with pre and post break slopes as $\alpha_{\rm pre}\simeq0.40$, $\alpha_{\rm post}\simeq-1.8$. The vertical green dotted line depicts the position of the break at a break frequency of $\simeq 4.2\times10^{-2}$ cycles. White noise/normalization breaks like these \citep{press1978} are generally not seen clearly in GeV PSDs, or are poorly constrained \citep{shah2025}. Nevertheless the post-break index is $\sim-2$ which is very similar to the general PSD indices observed for GeV light curves of bright blazars (\citealt{abdo2010,shah2025}, where indices $\sim -1.7$ were found).

\subsubsection{Skewness and Structural Entropy}

\begin{figure*}
    \centering
    \includegraphics[width=\linewidth]{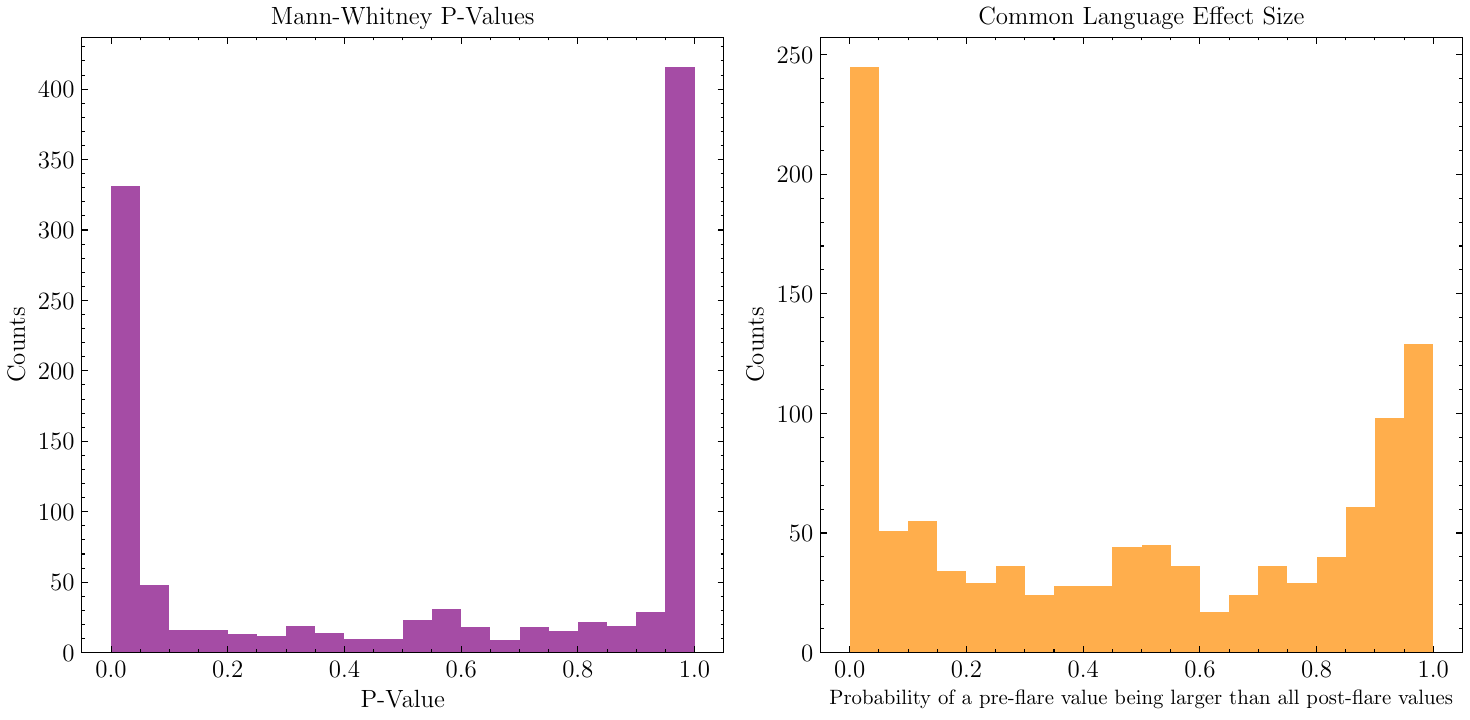}
    \caption{Left : A p-value histogram for the Mann Whitney U test for the decrease in skewness across the largest flare for  the 1500 simulated light curves. A bimodal distribution is evident, exhibiting two types of sources where skewness either decreases (low p-value) or increases (high p-value). Right : The corresponding common language effect size of the Mann Whitney Test, showing the number of light curves as a function of the probability where a random pair of pre and postflare skewness values has the preflare value larger than the postflare.}
    \label{fig:pvalue_skew_sim}
\end{figure*}

To understand the statistical properties of the light curves generated by our simulation, we run our simulation 1500 times to produce 1500 light curve samples. In addition to the rolling skewness diagnostic, we introduce an additional variable that measures the entropy of the system. The entropy can be grouped into two parts, which include the Shannon entropy \citep{shannon48} of the magnetic flux distribution ($\mathcal{S}_{\rm \psi}=-\Sigma p_\psi\ln p_\psi$) and a structural Shannon entropy of the $\delta_{\rm mj}$ distribution as $\mathcal{S}_\delta=-\Sigma p_\delta\ln p_\delta$. Both are equally weighted (although the latter is more dominant) and added to give the total entropy $\mathcal{S}=\mathcal{S}_\psi+S_\delta$. We then similarly use a rolling entropy window slid across the largest flare in each light curve for investigations of state transitions in the system. Like our analyses of real light curves we align all 1500 light curves centred on the flare peak and compute the rolling skewness and entropy. Note that when the window ran out of the points at its edges, they were padded with the last value of the window. Dividing the light curve into pre and postflare segments, we computed the Mann Whitney U statistics and the corresponding p-values for a reduction in skewness across the largest flare. Figure \ref{fig:pvalue_skew_sim} shows the auto-correlation corrected Mann Whitney U-test p-value histogram (left panel) and the corresponding effect size (right panel) as in Figure \ref{fig:pvalue_skew}. Using Hartigan's Dip Test \citep{hartigan1985}, unimodality in the p-value histogram is ruled out at a confidence exceeding thousands of $\sigma$. While the distribution is close to being bi-modal, the number of cases where skewness increases is $\sim 100$ higher, making the effect size larger at smaller probabilities in the right panel. Intermediate p-values are present, occupying a small fraction of the entire simulation. These are cases where skewness has not changed at a statistically significant level across the flare, similar to several cases in our observational sample.

\begin{figure*}
    \centering
    \hbox{\includegraphics[width=0.5\linewidth]{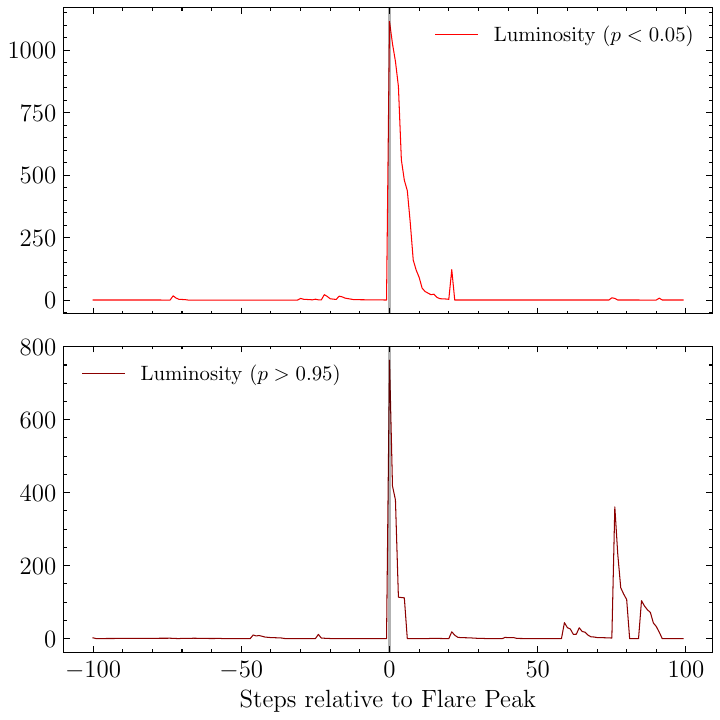}
    \includegraphics[width=0.5\linewidth]{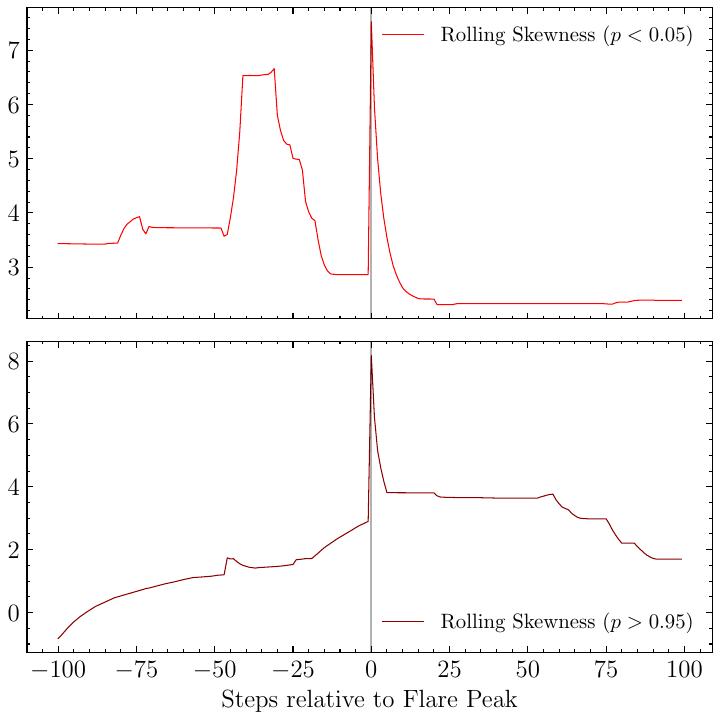}
    }
    \caption{An example of two light curves from our simulation from the lowest and highest parts of the p-value histogram. Both show large flare(s) and decrease in skewness around the flare, but only for the lighter red light curve (upper panel) one sees a persistent lower skewness than preflare levels. The darker red light curve (bottom panel), in contrast, has increased skewness after the flare.}
    \label{fig:sig_nonsig_sim}
\end{figure*}

Figure \ref{fig:sig_nonsig_sim} shows an example of a low and high p-value run, with corresponding luminosity and skewness plots as function of time. In the upper panel the skewness has fallen to a lower value than the preflare segment, and in the lower panel the skewness has increased across the flare.

\begin{figure*}
    \centering
    \includegraphics[width=\linewidth]{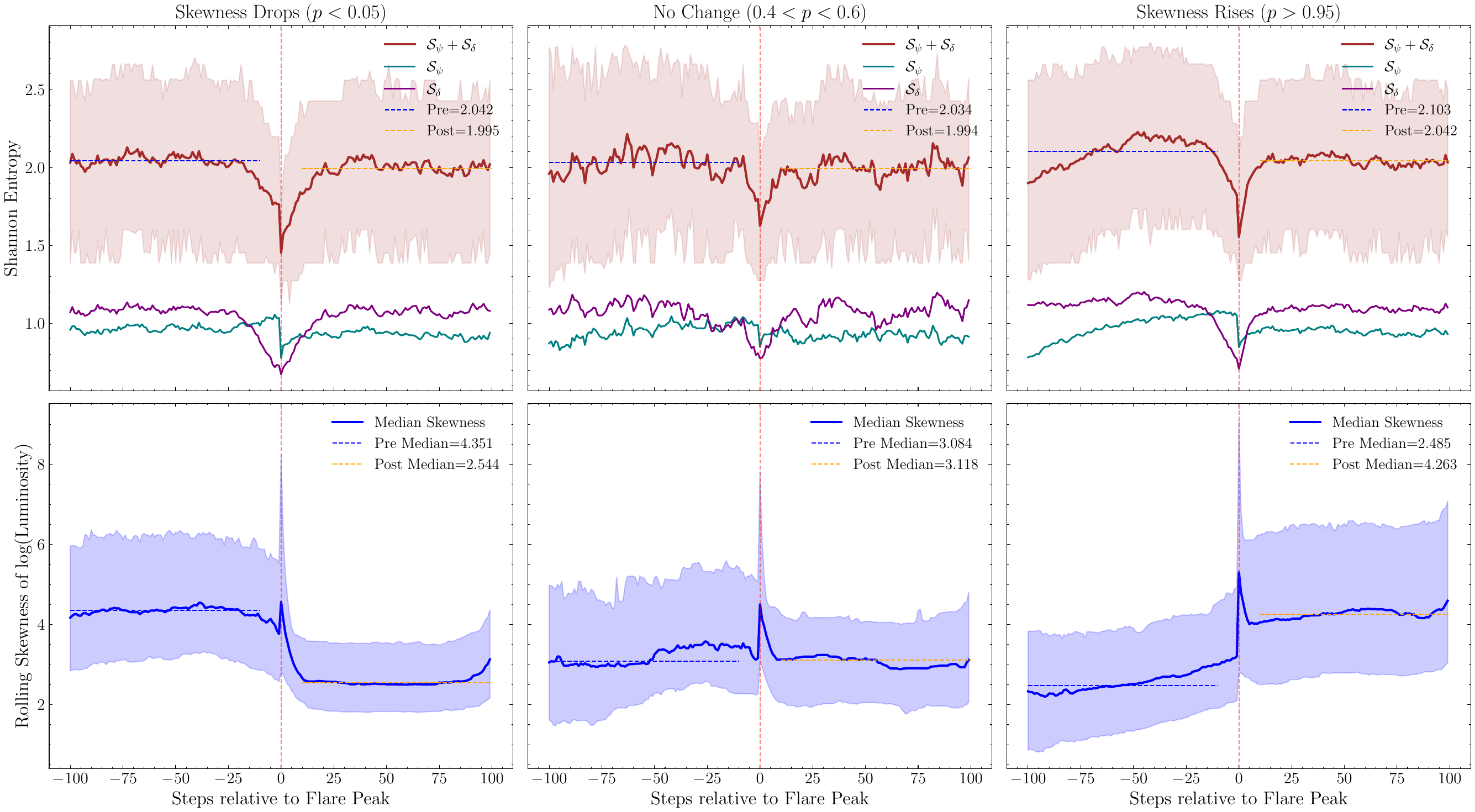}
    \caption{Figure shows the ensemble rolling Shannon entropy and skewness profiles of the 1500 light curves, categorized by Mann Whitney p-value, with left, middle and right panels showing that for $p<0.05$, $0.4<p<0.6$ and $p>0.95$ respectively. The skewness decreases, stays unchanged and increases in the left, middle and right panels respectively. However, the entropy is minimized at the flare and it reduces from the preflare values in both the edge cases of the p-values at a 3$\sigma$ level, while for the intermediate case it is 1$\sigma$, clearly showing an increase in order of the system as a result of loss in plasmoid number due to merging.}
    \label{fig:sim_skew}
\end{figure*}

We divide our runs into three categories on the basis of their Mann Whitney p-values as $p<0.05$, $0.4<p<0.6$ and $p>0.95$, which covers the entire gamut of the simulation. For each of these cases, we compute the ensemble entropy and the skewness profiles and plot in Figure \ref{fig:sim_skew}. The bottom panel shows the skewness profiles, with the bolder blue lines showing the median ensemble skewness. The spread is given by the 16th and 84th quartiles respectively. The observed behaviour is expected, where the skewness decreases by $\sim1-2$ in the $p<0.05$ case, and increases by a similar amount in the $p>0.95$ case. For the intermediate p-values, the skewness change is $\lesssim0.03$ and is not statistically significant. The upper panel shows the corresponding ensemble median of the total entropy (a solid dark red line) for each of these cases, and the geometric and magnetic flux components given in purple and teal respectively as solid lines. The spread is similarly computed with the 16th and 84th quartiles. For all the three cases, we see a sharp drop in entropy around the position of the largest flare. This is in accordance with a rapid increase in order of the system where the largest flare is produced by a rapid conglomeration of all plasmoids into a ``monster" plasmoid and its Doppler alignment at the same time. Both $\mathcal{S}_\psi$ and $\mathcal{S}_\delta$ show minima around the same point since the $\psi$ and $\delta$ distributions have peaked around single values during the largest flare. The entropy has more pronounced minima in the edge cases, while it is more noisy in the moderate cases, i.e. in the middle panel. The following rise in entropy is expected due to the continuous injection of plasmoids into the system, with the erstwhile monster plasmoid advected out of the system. In both the edge cases nevertheless, a reduction of entropy after a flare is evident (and at a $3\sigma$ level using a Mann-Whitney U Test; in the intermediate case the entropy is lower but only by $1\sigma$), i.e., the postflare entropy is lower than the preflare entropy. This is possible due to the loss in number of plasmoids, through escape and merging into the monster plasmoid. The large flare acts in a way to ``flush" the system of plasmoids, opening the way to the slow injection rates in our model, that are unable to recover the initial number of plasmoids $N_0=50$. The skewness, however, decreases, increases and stays the same depending on the case. The skewness quantifies long-tails in the observed log-luminosity distribution, and the distinction between the low and high p-value cases can be explained in the following way. In cases where the skewness decreases, it is possible the flare used up the only monster plasmoid in the system. This made the occurrence and magnitude of flares relative to the postflare luminosity baseline go down. In cases where the skewness increases, there are at least few more plasmoids remaining after the flare. These, along with the continuous injection, can merge and produce flares \textit{still} on top of a relatively stabler background than the preflare (since the number of plasmoids have gone down and hence mergers) making the skewness go up. This is because the skewness is not a direct proxy for the flaring rate, but it instead measures the \textit{amount of departure} from the baseline luminosity level. We also note that the geometric entropy $\mathcal{S}_\delta$ recovers its previous state due to new plasmoid injection, but the magnetic flux entropy $\mathcal{S}_\psi$ reduces across the flare regardless. This is due to the fact that injection introduces new plasmoids that contain only seed flux and reaching the preflare flux baseline is hence much more difficult than the Doppler factor. Hence in the $p<0.05$ cases, the only large plasmoid was advected out of the system, reducing flaring rapidly, and hence decreasing skewness. In the $p>0.95$ case, few other monster plasmoids remained on top of a \textit{now-steadier} background luminosity, thereby making the skewness go up across the flare. In the intermediate case, the monster flare has been unable to increase order in the system at a statistically significant level. The injection rate is relatively strong enough compared to escape in these systems that the flaring rates are effectively maintained and disorder is sustained. This shows up in the skewness where departures from the mean luminosity level are minimal.

The above discussion naturally ties in with our observational results in Section \ref{sec:obs}. The observational sample is not statistically complete, spread through the entire p-value range, with larger counts at the edges. 1000 random choices of 18 realizations from our simulation from Figure \ref{fig:pvalue_skew_sim} and thereafter comparison with the observed p-value histogram in Figure \ref{fig:pvalue_skew} using a Kolmogorov-Smirnov Test \citep{berger2014} produce a KS test p-value distribution where 96\% of the p-values lie at $p_{\rm KS}>0.05$. While noting the statistical weakness of the observational sample, this implies the simulated Mann Whitney U p-value distribution cannot be distinguished from the observed MW p-value distribution at a statistically significant level. While it might be argued that any other model could also possibly produce the observed p-value distribution after downsampling to 18 samples, such a test is not possible without a larger observational sample. A larger and statistically complete sample of FSRQs is out of scope of this paper and is the focus of a future work.

We further note that the median skewness values produced by our simulation are consistently few units higher than those calculated from the \textit{Fermi}-LAT observations, $\mathbb{S}_{\rm model}-\mathbb{S}_{\rm obs}\sim3-4$. The offset is lower for $\Delta\mathbb{S}$, with $\Delta\mathbb{S}_{\rm model}-\Delta\mathbb{S}_{\rm obs}\sim1$. These offsets are not unexpected as our simulation lacks the Poisson noise inherent in real photon counting data like Fermi-LAT as well as the noisy background emission from the jet bulk (which we only modelled as a baseline flux), that naturally reduces higher-order moments, and hence the skewness. However, the physical significance of our result lies in the direction of change in the skewness $\Delta \mathbb{S}$ across the flare, and the Mann Whitney U p-value distribution for the entire population, both of which are recovered self-consistently.

\subsubsection{Test of Stationarity of The Power Spectral Density}

\begin{figure*}
    \centering
    \includegraphics[width=0.75\linewidth]{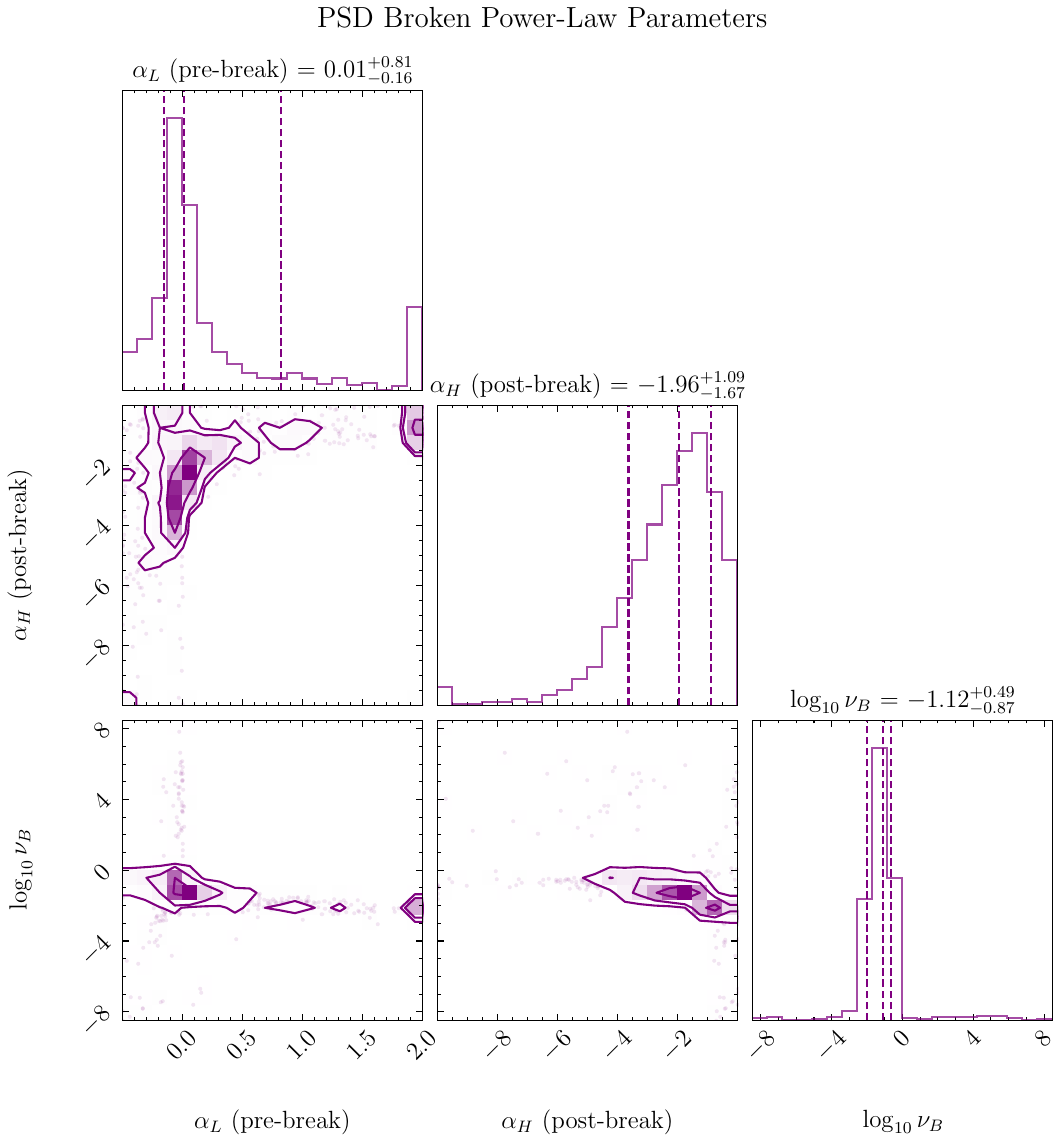}
    \caption{Corner plot of all best-fit broken power law parameters of the power spectral density of our simulated light curves. The prebreak slope is white noise $\sim0.01$, the break frequency peaks at $\sim0.07$ cycles/s (which is close to the inverse of the cooling timescale in our simulation) and the postbreak is clearly red noise at $\sim -2.0$, characteristic of historically observed blazar variability.}
    \label{fig:total_psd_slopes}
\end{figure*}

In order to understand the evolution of the power spectral density across the flare and the total PSD of the light curve, we fit a typical broken-power law PSD to the pre and postflare segments, and the total light curve. Figure \ref{fig:total_psd_slopes} shows the corner plot of all the fitted break frequencies (in simulation units), the pre-break slope and the post-break slopes of the ensemble of simulated light curves.

The median pre-break slope is $\sim0.01$, consistent with white noise, the post-break slope is $\sim-2.0$, providing a direct recovery of red noise. The large error bar for the prebreak slopes is due to the outliers at the higher end of the histogram, which are a result of poor fitting. The break frequency is $\sim10^{{-1.12}^{+0.49}_{-0.87}}=0.07^{+0.18}_{-0.06}$, ranging between $0.01$ and $0.25$. The cooling factor is $\kappa=0.05$ in our simulation setup. Given $\delta_{\rm jet}=3.0$, if the break frequency represents the cooling factor/timescale, then it should be $\sim0.05\delta_{\rm jet}=0.15$, which is within $1\sigma$ of the median break frequency. There is no other relevant timescale in our system that is deterministic, since plasmoid merging and state changes can happen anytime due to it being a stochastic process. We must also note that a break is not observed in all the light curves, and around $\sim30\%$ could not either be fit with a broken power law and instead the post-break slope represented a single power-law fit in those cases or could not be fit at all. Hence while the cooling timescale might be a plausible explanation for the break timescale, combinations of the injection-merger-escape rates might provide an explanation for the break timescale too, which might include a timescale in the system after which either of the three rates dominate. However, since for a physically relevant range in initial parameters (Appendix \ref{app:A}), the median of the PSD break frequency histogram shifts negligibly, it is more likely that the break timescale is directly representative of the cooling timescale with the spread being the result of a purely stochastic simulation, which would depend on the relevant parameters too. A further detailed parametric study is out of scope of this work.

\begin{figure*}
    \centering
    \includegraphics[width=0.75\linewidth]{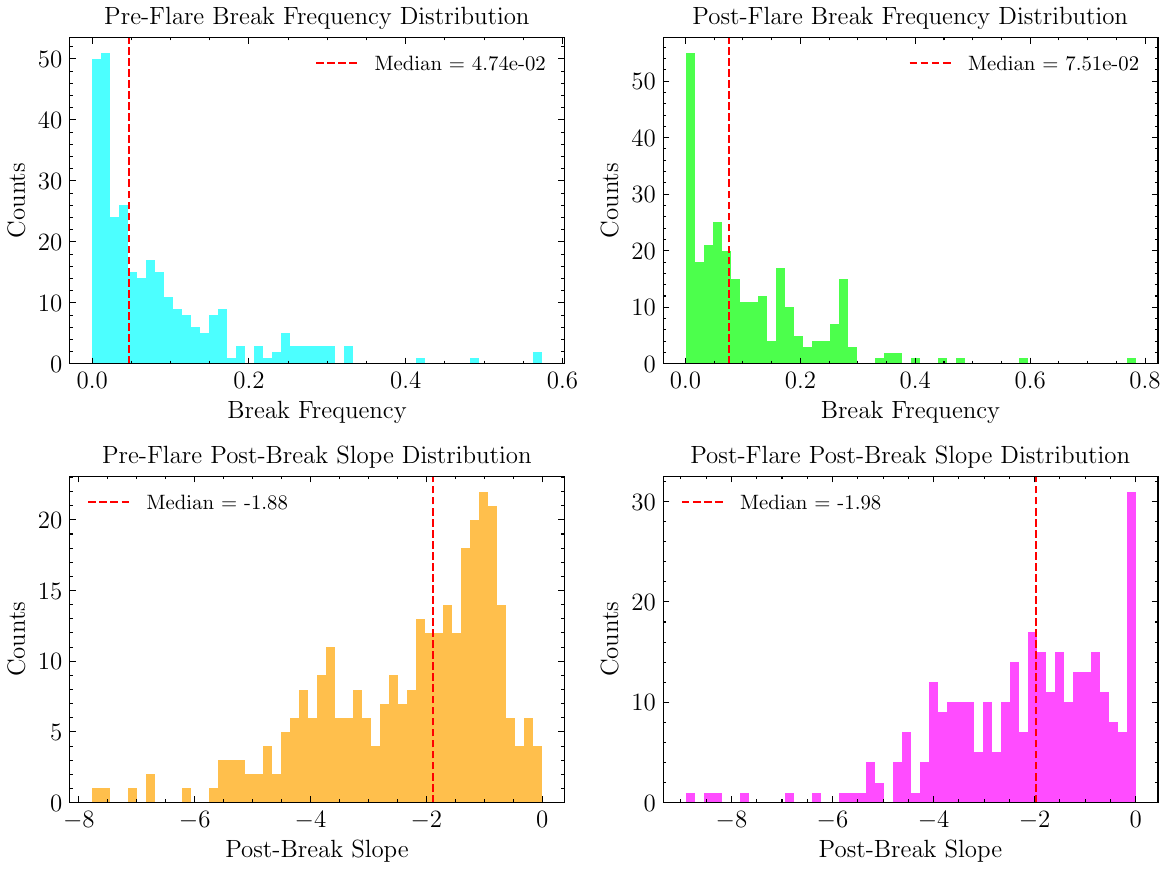}
    \caption{Histograms of the best-fit PSD parameters (break frequencies and postbreak slopes) for both the preflare and postflare segments in our simulated light curves. Neither the break frequency nor the postbreak slope medians are significantly different, being close to the cooling timescale in our simulation and consistent with red noise respectively.}
    \label{fig:pre_post_psd_hist}
\end{figure*}

We continue the above analysis for a test of stationarity of the power spectral density across the largest flare in each run. To understand if the pre and postflare segments show similar PSD properties, we compute the PSDs and plotted the histograms of the break frequencies and the postbreak slopes in Figure \ref{fig:pre_post_psd_hist}. It is evident from the figure that there is no distinct difference neither in pre or postflare PSDs slopes or the break frequencies, or are within error bars. The blazar producing red noise before and after the flares suggests our statistical model self-consistently produces it, very similar to what has been observationally found.

However, in order to test if the pre and postflare histograms are fundamentally different, we ran a two-sample Kolmogorov-Smirnov test \citep{berger2014} for the both the break frequency histogram and the postbreak slopes. The former pair is \textit{different} at a p-value of $\sim 0.03$ and the latter pair has a p-value $\sim 0.40$. Hence the break frequency histogram is fundamentally different at a $\sim 2\sigma$ while the postbreak slope histogram changes through the flare at a $\sim 1 \sigma$ level. This implies that while the median values have not changed through the flare, some feature of the system might have changed that resulted in the underlying distribution for the PSD parameters to change. The macroscopic PSD behaviour may not hence change due to large flares, but it envelopes microscopic changes in variability due to large flares which the global PSD cannot capture.

The above simulation was for a given set of parameters. We show in Appendix \ref{app:A} that for a wide range of physical parameters the observed trends in the skewness and the power spectral density are modified negligibly, implying the observational similarities can be  generalized and are not just a result of a very finely tuned initial condition of parameters defined in Table \ref{tab:pars}. However, even without testing one can easily discern that a large value of $S_{\rm base}\gtrsim1.0$ will introduce too many plasmoids into the system and the observed luminosity profile will have many GeV flares, preventing a state transition. Too low of an injection rate will preclude this entire analysis and all similarities with observed blazar variability, as also shown in Appendix \ref{app:A}. Since it is not possible a priori to perfectly decide on the values of these free parameters, it is only through similarities with observations one can provide a prior for the model. We hence come to the conclusion that this state transition to a stabler state is a robust feature of the reconnection-driven stochastic process. 

All of our simulations were produced in arbitrary units, scaled only by $L/d_ic_A$. Given $c_A\simeq 1$, our timescales can be transformed to observable real units through appropriate scaling of the current sheet length $L/d_i$. \cite{giannios13} mentions a minimum reconnection length scale to around $\sim 0.1-1$ pc using observational data. If the ion skin depth equals unity still, this would make the timescales $0.3-3$ years, covering $60-600$ years in our time-window of 200 steps as in Figure \ref{fig:sig_nonsig_sim}. However, this is way larger than the observational durations and further the ion skin depth is difficult to constrain. Since real jets can clearly show this kind of a state transition and we use \textit{that} as a starting pillar, the sheet lengths $L/d_i$ should be close to $0.05-0.1$ pc to match a rough 20 year window with 200 time steps in our simulation. 

\section{Discussion}
\label{sec:disc}
Magnetic reconnection as a proposed model for blazar variability has existed in the literature for more than a decade now, especially after Fermi-LAT started monitoring flaring blazars. The closest interpretation to blazar variability has been \cite{giannios13} where the author demonstrated that plasmoids grow to form a monster plasmoid that is responsible for fast GeV-TeV flares. The slower envelope of emission would come from all the smaller plasmoids. The treatment was mostly analytical and it relied on plasmoid size growth instead of \textit{mergers}, like our work here. Hence, without a statistical/full radiative particle-in-cell (PIC) model (which was not possible) fully reproducing observational trends self-consistently was not possible. Further, Fermi-LAT had only observed for five years in 2013, preventing any long-term statistical light curve analyses. Various other works on reconnection have mostly tried to compare the efficiency of shock acceleration with reconnection (e.g., \citealt{sironi15} or \citealt{petrop19} and relevant references therein) for blazar emission. Our model assumes reconnection as the primary source of energy for blazar emission and proceeds with the statistical plasmoid merger model of \cite{fermo2010} and stochastically solves the evolution of the plasmoid magnetic flux distribution, in turn allowing to capture the full stochastic nature of blazar variability as demonstrated in multiple figures throughout the text.

In this paper, we followed up on a previous publication on CTA 102 \citep{royc25} to investigate in detail if the putative state transition in CTA 102 is universal or only a source-specific effect. Our simulation results have been able to effectively capture all of their statistical nature, making it a viable model for the origin of $\gamma$-ray variability. The sensitivity of the key results, the nature of the p-value distribution, and PSD parameters, to a wide range of parameters was explored in Appendix \ref{app:A} and were found to be mostly robust to parameter changes as long as they are physical. Multi-wavelength flares are often observed though, but they many a time lack the intensity and variability timescales of GeV emission, which possibly arise from regions much smaller than the light crossing time of the bulk jet. In contrast, in the context of our model, optical or X-ray flares for example could be possibly modelled using larger cooling timescales and with a better treatment of the magnetic field evolution during mergers. Further, the emitting regions are generally larger for lower energy emission, which, for a given magnetic field, would have a reduced mass and magnetic energy density. As per model assumptions, we further note that the prescription of direct equivalence between dissipated energy and radiative luminosity is rather simplified. It assumes a fraction of the dissipated energy ($f_{\rm gain}\psi^2$) fully goes into GeV radiation, ignoring the microphysics of the heating treatment. A possible next step would be normalizing a non-thermal electronic energy distribution by the energy dissipated, in proper units. However, the EED index would need to be inspired from Particle-in-Cell (PIC) simulations, without a direct connection to amount of energy dissipated. This is a plan for future work, where a multi-wavelength prediction could be assessed. Any further movement into rigor would require a full PIC treatment that heats particles self-consistently.


The behaviour of the Shannon entropy in Figures \ref{fig:sim_skew} indicates an increase in order in the system due to plasmoid mergers. It is minimized at the flare in all realizations and decreases to a lower value in the edge cases of the Mann Whitney p-value histogram, characterizing transition to a more ordered state. The skewness spikes around the flare, which can be qualitatively likened to a discontinuity, but it does not meet all criteria to be classified as an order parameter for a non-equilibrium phase transition. The monster flare increases order in the system and the skewness attains a persistent higher and lower value in most of the simulation runs, marking a state transition. This is unrelated to the actual thermodynamics of the jet plasma itself and is akin to general structural phase transitions seen in a wide variety of physical situations \citep[e.g.,][]{Haken1989,hinrichsen2000,vellela2009, bouchet2012,caroll2020} where phase transitions are a hallmark of stochastic thermodynamics in so-called non-equilibrium steady states (NESS) \citep[see e.g.,][]{seifert12}. The almost-discontinuous change in the skewness, and the reduction in entropy across the flare thereafter indicate that the blazar has possibly undergone a driven state-transition between two non-equilibrium steady states, since the process does not meet all criteria required for a phase transition. This is in marked contrast with the thermodynamics of the jet plasma itself, since it generally cannot be studied using standard equilibrium thermodynamics, as the majority of observed radiation comes from non-thermal particles. The thermal jet gas will have its own thermodynamics, but the presence of strong shocks and instabilities renders that assumption inaccurate and hence make analysis more difficult. Stochasticity is a well known phenomenon of blazar jet variability and this study provides a framework to investigate the same through a stochastic plasmoid merger model inside a jet. 



\section{Conclusions}
\label{sec:conc}

1. The source-specific work of \cite{royc25} on CTA 102 was extended to 17 more FSRQs in this paper. Based on the statistical characteristics of these FSRQs, they could be divided into three categories. One, where the skewness of the logarithmic flux distribution reduces after a large flare. Second, where the skewness increases instead. Third, where the skewness change is not statistically significant. This results in a spread-out distribution of p-values (although with higher values at the edges) from a Mann Whitney U Test conducted between the preflare and postflare skewness distribution of each source.

2. For sources where the skewness has decreased, they fall into the category where a large flare has significantly shifted the system to a steadier state, with less flaring, both in rate and magnitude. Where the skewness has increased instead, the system has reached a state where flares are more dominant than preflare and not necessarily more frequent. 



3. We used a statistical model of plasmoids in a current layer derived from \cite{fermo2010} that self-consistently takes into account plasmoid injection, escape and merging. When plasmoids merge, their field lines reconnect and release energy. We derive the luminosity as a proxy for the total internal energy of the system (for all plasmoids) and incorporate phenomenological cooling for the sake of preventing a full radiative treatment. We ensure that their orientations remain isotropic in the frame of the jet, mimicking the minijets-in-a-jet model of \cite{biteau12}. The observed luminosity is hence produced through Lorentz boosting, once in the rest-frame of the jet and then to the lab frame for the bulk jet. Our results produce large flares when a ``monster" plasmoid lines up towards our line of sight \citep[e.g.,]{giannios13}. 

4. For a total of 50 initial plasmoids and relevant simulation parameters, we find a clear bi-modal distribution of p-values for reduction in skewness, with a small fraction of intermediate cases where skewness change is undetermined. Downsampling the 1500 runs to 18 runs to compare with observations, a KS-test results in a statistically non-significant distinguishability between the observed and downsampled p-value histograms. This further validates the assumptions in our model, since the types of FSRQs in our observational sample can be effectively captured by the simulation. Future work would include a statistically complete sample of FSRQs.

5. We compute the Shannon entropy as a sum of the Shannon entropy of the magnetic flux distribution of the plasmoids and their Doppler factor distribution. We find the ensemble entropy is minimized around the largest flare, then decays to a lower value than the preflare in the edge p-value cases, with undetermined changes in the intermediate category. This is a direct signature of increase in order after a flare, due to the reduction in number of plasmoids and there remaining only a few large monster plasmoids. This implies a state transition between two non-equilibrium steady states \citep{seifert12} in the blazar is possibly occurring here.

6. We computed the power spectral densities of the 1500 realizations and we find that they can be represented best by a broken power law. The prebreak slopes are mostly white noise but the postbreak is consistently red noise, akin to the historically understood red noise behaviour of blazar variability. The break frequency is found to be close to the cooling timescale in our runs.

7. We further analyzed the PSD characteristics of the preflare and postflare segments of each simulated light curve. The best-fit slopes and break frequency have mostly remained unchanged, with a hint of a change in the underlying distribution through the flare.

8. Our stochastic simulation of the \cite{fermo2010} plasmoid merger model using Gillespie's algorithm has successfully reproduced several observable characteristics of blazar GeV variability, especially the nature of their flaring, the typical state transitions via skewness in our sample of FSRQs and the nature of the power-spectral density. 

9. This is a first direct observational study of possible state transitions in FSRQs occurring when a monster plasmoid forms in the system. However, this work is only limited to understanding GeV flaring and the corresponding changes in the blazar system. Future work would involve incorporation of a full multi-wavelength radiative treatment and a detailed mathematical analysis of the entropy.

\section{Acknowledgments}

We acknowledge the support of the Department of Atomic Energy, Government of India, under the project 12-R\&D-TFR-5.02-0700 and the anonymous referee whose suggestions strongly improved the manuscript.

\appendix

\section{Results of the simulation run with variations in six physical parameters}
\label{app:A}

\begin{figure}
    \centering
    \includegraphics[width=\linewidth]{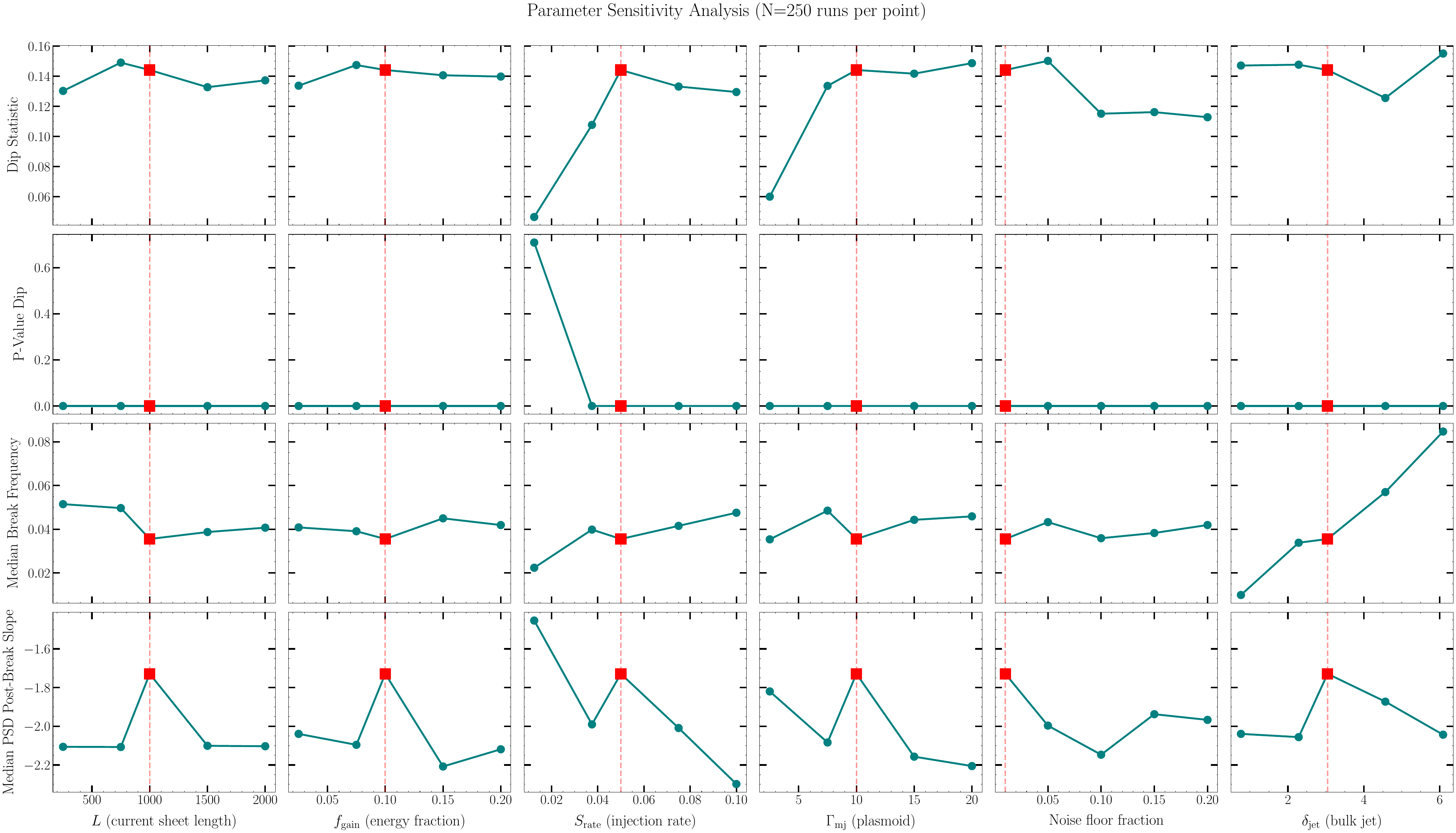}
    \caption{Figure shows the sensitivity of four statistical outputs as function of six physical parameters in the model. The six parameters consist of the sheet length $L$, energization fraction $f_{\rm gain}$, injection rate $S_{\rm base}$, the fraction of peak luminosity as baseline bulk emission, the Lorentz factor of the plasmoids $\Gamma_{\rm mj}$ and the Doppler factor of the bulk jet $\delta_{\rm jet}$. The simulation's output is represented by four statistical parameters, namely the Hartigan's Dip Statistic and the corresponding p-value for the Mann Whitney p-value histogram in addition to the median break frequency and post-break slope in the power spectral densities. The results are negligibly sensitive to the input parameters, with no clear trend, barring the case of a very low injection rate that removes bimodality of the Mann Whitney p-value histogram (Dip p-value close to 1) and increase in $\delta_{\rm jet}$ that naturally increases the PSD break frequency. The red point and line in each sub-figure denote the fiducial parameter set chosen in the main text in Table \ref{tab:pars}.}
    \label{fig:sens}
\end{figure}

Figure \ref{fig:sens} shows the effect of varying six important parameters of our model in Table \ref{tab:pars}, namely the sheet length $L$, energization fraction $f_{\rm gain}$, injection rate $S_{\rm base}$, the fraction of peak luminosity as baseline bulk emission, the Lorentz factor of the plasmoids $\Gamma_{\rm mj}$ and the Doppler factor of the bulk jet $\delta_{\rm jet}$ on the Hartigan's Dip Statistic and the corresponding p-value for the Mann Whitney p-value histogram in addition to the median break frequency and post-break slope in the power spectral densities. Except $\delta_{\rm jet}$, all other parameters were reduced or increased by at least $\sim2-3$ times their fiducial values (those that were used in the main results). We find that the bimodality in the Mann Whitney U histograms remain all throughout, except where the injection rate is too low to sustain the variability (second panel from the top, third figure from the left). In all cases, the median break frequency has remained fairly stable and has expectedly increased due to increase in $\delta_{\rm jet}$ that shortens all timescales due to relativistic aberration. The median post-break slopes show a spread mainly between $-(1.7-2.2)$, which is fully consistent with red noise behaviour. Our results are also stable across varying sheet lengths, implying larger or smaller systems that undergo stochastic flaring could possibly be another avenue for this model, in addition to accounting for the uncertainty in sheet lengths for blazar jets itself.

\bibliography{version2}{}

@ARTICLE{royc25,
       author = {{Roychowdhury}, Agniva},
        title = "{Minijets and Broken Stationarity in a Blazar: Novel Insights into the Origin of {\ensuremath{\gamma}}-Ray Variability in CTA 102}",
      journal = {\apj},
         year = 2026,
        month = feb,
       volume = {997},
       number = {2},
          eid = {184},
        pages = {184},
          doi = {10.3847/1538-4357/ae3140},
archivePrefix = {arXiv},
       eprint = {2512.21240},
 primaryClass = {astro-ph.HE},
       adsurl = {https://ui.adsabs.harvard.edu/abs/2026ApJ...997..184R}
}

@article{hartigan1985,
  author = {Hartigan, J. A. and Hartigan, P. M.},
  title = {The Dip Test of Unimodality},
  journal = {The Annals of Statistics},
  year = {1985},
  volume = {13},
  number = {1},
  pages = {70--84},
  doi = {10.1214/aos/1176346577},
  publisher = {Institute of Mathematical Statistics}
}

@article{huang2010,
    author = {Huang, Yi-Min and Bhattacharjee, A.},
    title = {Scaling laws of resistive magnetohydrodynamic reconnection in the high-Lundquist-number, plasmoid-unstable regime},
    journal = {Physics of Plasmas},
    volume = {17},
    number = {6},
    pages = {062104},
    year = {2010},
    month = {06},
    issn = {1070-664X},
    doi = {10.1063/1.3420208},
    url = {https://doi.org/10.1063/1.3420208},
    eprint = {https://pubs.aip.org/aip/pop/article-pdf/doi/10.1063/1.3420208/16029914/062104_1_online.pdf},
}

@incollection{brown2005,
title = {Course 14 - Theory of Point Processes for Neural Systems},
editor = {C.C. Chow and B. Gutkin and D. Hansel and C. Meunier and J. Dalibard},
series = {Les Houches},
publisher = {Elsevier},
volume = {80},
pages = {691-727},
year = {2005},
booktitle = {Methods and Models in Neurophysics},
issn = {0924-8099},
doi = {https://doi.org/10.1016/S0924-8099(05)80020-4},
url = {https://www.sciencedirect.com/science/article/pii/S0924809905800204},
author = {Emery N. Brown}
}

@ARTICLE{ogata81,
  author={Ogata, Y.},
  journal={IEEE Transactions on Information Theory}, 
  title={On Lewis' simulation method for point processes}, 
  year={1981},
  volume={27},
  number={1},
  pages={23-31},
  doi={10.1109/TIT.1981.1056305}}

@article{davies99,
doi = {10.1088/0305-4470/32/44/311},
url = {https://doi.org/10.1088/0305-4470/32/44/311},
year = {1999},
month = {nov},
publisher = {},
volume = {32},
number = {44},
pages = {7745},
author = {Susan C Davies and John R King and Jonathan A D Wattis},
title = {The 
Smoluchowski 
coagulation equations with continuous injection},
journal = {Journal of Physics A: Mathematical and General}
}

@article{vellela2009,
  author    = {Vellela, M. and Qian, H.},
  title     = {Stochastic dynamics and non-equilibrium thermodynamics of a bistable chemical system: the {Schlögl} model revisited},
  journal   = {Journal of the Royal Society Interface},
  year      = {2009},
  volume    = {6},
  number    = {39},
  pages     = {925--940},
  month     = {Oct},
  doi       = {10.1098/rsif.2008.0476},
  note      = {Epub 2008 Dec 18},
  pmid      = {19095615},
  pmcid     = {PMC2838355}
}

@article{hinrichsen2000,
author = {Haye Hinrichsen},
title = {Non-equilibrium critical phenomena and phase transitions into absorbing states},
journal = {Advances in Physics},
volume = {49},
number = {7},
pages = {815--958},
year = {2000},
publisher = {Taylor \& Francis},
doi = {10.1080/00018730050198152},


URL = { 
    
        https://doi.org/10.1080/00018730050198152
    
    

},
eprint = { 
    
        https://doi.org/10.1080/00018730050198152
    
    

}

}

@article{Haken1989,
doi = {10.1088/0034-4885/52/5/001},
url = {https://doi.org/10.1088/0034-4885/52/5/001},
year = {1989},
month = {may},
publisher = {},
volume = {52},
number = {5},
pages = {515},
author = {H Haken},
title = {Synergetics: an overview},
journal = {Reports on Progress in Physics}
}

@article{caroll2020,
title = {Geometry of quantum phase transitions},
journal = {Physics Reports},
volume = {838},
pages = {1-72},
year = {2020},
note = {Geometry of quantum phase transitions},
issn = {0370-1573},
doi = {https://doi.org/10.1016/j.physrep.2019.11.002},
url = {https://www.sciencedirect.com/science/article/pii/S0370157319303655},
author = {Angelo Carollo and Davide Valenti and Bernardo Spagnolo}
}

@article{seifert12,
doi = {10.1088/0034-4885/75/12/126001},
url = {https://doi.org/10.1088/0034-4885/75/12/126001},
year = {2012},
month = {nov},
publisher = {IOP Publishing},
volume = {75},
number = {12},
pages = {126001},
author = {Seifert, Udo},
title = {Stochastic thermodynamics, fluctuation theorems and molecular machines},
journal = {Reports on Progress in Physics}
}

@article{bouchet2012,
title = {Statistical mechanics of two-dimensional and geophysical flows},
journal = {Physics Reports},
volume = {515},
number = {5},
pages = {227-295},
year = {2012},
note = {Statistical mechanics of two-dimensional and geophysical flows},
issn = {0370-1573},
doi = {https://doi.org/10.1016/j.physrep.2012.02.001},
url = {https://www.sciencedirect.com/science/article/pii/S0370157312000518},
author = {Freddy Bouchet and Antoine Venaille}
}

@ARTICLE{mukherjee19,
       author = {{Mukherjee}, Sagnick and {Mitra}, Kaustav and {Chatterjee}, Ritaban},
        title = "{The accretion disc-jet connection in blazars}",
      journal = {\mnras},
         year = 2019,
        month = jun,
       volume = {486},
       number = {2},
        pages = {1672-1680},
          doi = {10.1093/mnras/stz858},
archivePrefix = {arXiv},
       eprint = {1904.03740},
 primaryClass = {astro-ph.HE},
       adsurl = {https://ui.adsabs.harvard.edu/abs/2019MNRAS.486.1672M}
}

@ARTICLE{sironi25,
       author = {{Sironi}, Lorenzo and {Uzdensky}, Dmitri A. and {Giannios}, Dimitrios},
        title = "{Relativistic Magnetic Reconnection in Astrophysical Plasmas: A Powerful Mechanism of Nonthermal Emission}",
      journal = {\araa},
         year = 2025,
        month = aug,
       volume = {63},
       number = {1},
        pages = {127-178},
          doi = {10.1146/annurev-astro-020325-115713},
archivePrefix = {arXiv},
       eprint = {2506.02101},
 primaryClass = {astro-ph.HE},
       adsurl = {https://ui.adsabs.harvard.edu/abs/2025ARA&A..63..127S}
}

@ARTICLE{petrop19,
       author = {{Petropoulou}, Maria and {Sironi}, Lorenzo and {Spitkovsky}, Anatoly and {Giannios}, Dimitrios},
        title = "{Relativistic Magnetic Reconnection in Electron-Positron-Proton Plasmas: Implications for Jets of Active Galactic Nuclei}",
      journal = {\apj},
         year = 2019,
        month = jul,
       volume = {880},
       number = {1},
          eid = {37},
        pages = {37},
          doi = {10.3847/1538-4357/ab287a},
archivePrefix = {arXiv},
       eprint = {1906.03297},
 primaryClass = {astro-ph.HE},
       adsurl = {https://ui.adsabs.harvard.edu/abs/2019ApJ...880...37P}
}

@ARTICLE{das26,
       author = {{Das}, Chandan Kumar and {Vaidya}, Bhargav and {Shukla}, Amit and {Mattia}, Giancarlo and {Mannheim}, Karl},
        title = "{Role of Magnetic Reconnection in Blazar Variability Using Numerical Simulation}",
      journal = {\apj},
         year = 2026,
        month = jan,
       volume = {996},
       number = {1},
          eid = {53},
        pages = {53},
          doi = {10.3847/1538-4357/ae2330},
archivePrefix = {arXiv},
       eprint = {2511.19605},
 primaryClass = {astro-ph.HE},
       adsurl = {https://ui.adsabs.harvard.edu/abs/2026ApJ...996...53D}
}

@ARTICLE{petrop18,
       author = {{Petropoulou}, M. and {Christie}, I.~M. and {Sironi}, L. and {Giannios}, D.},
        title = "{Plasmoid statistics in relativistic magnetic reconnection}",
      journal = {\mnras},
         year = 2018,
        month = apr,
       volume = {475},
       number = {3},
        pages = {3797-3812},
          doi = {10.1093/mnras/sty033},
archivePrefix = {arXiv},
       eprint = {1710.00724},
 primaryClass = {astro-ph.HE},
       adsurl = {https://ui.adsabs.harvard.edu/abs/2018MNRAS.475.3797P}
}

@ARTICLE{sironi15,
       author = {{Sironi}, Lorenzo and {Petropoulou}, Maria and {Giannios}, Dimitrios},
        title = "{Relativistic jets shine through shocks or magnetic reconnection?}",
      journal = {\mnras},
         year = 2015,
        month = jun,
       volume = {450},
       number = {1},
        pages = {183-191},
          doi = {10.1093/mnras/stv641},
archivePrefix = {arXiv},
       eprint = {1502.01021},
 primaryClass = {astro-ph.HE},
       adsurl = {https://ui.adsabs.harvard.edu/abs/2015MNRAS.450..183S}
}

@inbook{berger2014,
author = {Berger, Vance W. and Zhou, YanYan},
publisher = {John Wiley \& Sons, Ltd},
isbn = {9781118445112},
title = {Kolmogorov–Smirnov Test: Overview},
booktitle = {Wiley StatsRef: Statistics Reference Online},
doi = {https://doi.org/10.1002/9781118445112.stat06558},
url = {https://onlinelibrary.wiley.com/doi/abs/10.1002/9781118445112.stat06558},
eprint = {https://onlinelibrary.wiley.com/doi/pdf/10.1002/9781118445112.stat06558},
year = {2014},
}

@ARTICLE{shannon48,
       author = {{Shannon}, C.~E.},
        title = "{A mathematical theory of communication}",
      journal = {Bell Labs Technical Journal},
         year = 1948,
        month = jul,
       volume = {27},
       number = {3},
        pages = {379-423},
          doi = {10.1002/j.1538-7305.1948.tb01338.x},
       adsurl = {https://ui.adsabs.harvard.edu/abs/1948BSTJ...27..379S}
}

@ARTICLE{abdo2010,
       author = {{Abdo}, A.~A. and {Ackermann}, M. and {Ajello}, M. and {Antolini}, E. and {Baldini}, L. and {Ballet}, J. and {Barbiellini}, G. and {Bastieri}, D. and {Bechtol}, K. and {Bellazzini}, R. and {Berenji}, B. and {Blandford}, R.~D. and {Bloom}, E.~D. and {Bonamente}, E. and {Borgland}, A.~W. and {Bouvier}, A. and {Bregeon}, J. and {Brez}, A. and {Brigida}, M. and {Bruel}, P. and {Buehler}, R. and {Burnett}, T.~H. and {Buson}, S. and {Caliandro}, G.~A. and {Cameron}, R.~A. and {Caraveo}, P.~A. and {Carrigan}, S. and {Casandjian}, J.~M. and {Cavazzuti}, E. and {Cecchi}, C. and {{\c{C}}elik}, {\"O}. and {Chekhtman}, A. and {Cheung}, C.~C. and {Chiang}, J. and {Ciprini}, S. and {Claus}, R. and {Cohen-Tanugi}, J. and {Cominsky}, L.~R. and {Conrad}, J. and {Costamante}, L. and {Cutini}, S. and {Dermer}, C.~D. and {de Angelis}, A. and {de Palma}, F. and {Silva}, E. do Couto e. and {Drell}, P.~S. and {Dubois}, R. and {Dumora}, D. and {Farnier}, C. and {Favuzzi}, C. and {Fegan}, S.~J. and {Focke}, W.~B. and {Fortin}, P. and {Frailis}, M. and {Fukazawa}, Y. and {Funk}, S. and {Fusco}, P. and {Gargano}, F. and {Gasparrini}, D. and {Gehrels}, N. and {Germani}, S. and {Giebels}, B. and {Giglietto}, N. and {Giommi}, P. and {Giordano}, F. and {Glanzman}, T. and {Godfrey}, G. and {Grenier}, I.~A. and {Grondin}, M.-H. and {Grove}, J.~E. and {Guiriec}, S. and {Hadasch}, D. and {Hayashida}, M. and {Hays}, E. and {Healey}, S.~E. and {Horan}, D. and {Hughes}, R.~E. and {Itoh}, R. and {J{\'o}hannesson}, G. and {Johnson}, A.~S. and {Johnson}, W.~N. and {Kamae}, T. and {Katagiri}, H. and {Kataoka}, J. and {Kawai}, N. and {Kn{\"o}dlseder}, J. and {Kuss}, M. and {Lande}, J. and {Larsson}, S. and {Latronico}, L. and {Lemoine-Goumard}, M. and {Longo}, F. and {Loparco}, F. and {Lott}, B. and {Lovellette}, M.~N. and {Lubrano}, P. and {Madejski}, G.~M. and {Makeev}, A. and {Massaro}, E. and {Mazziotta}, M.~N. and {McEnery}, J.~E. and {Michelson}, P.~F. and {Mitthumsiri}, W. and {Mizuno}, T. and {Moiseev}, A.~A. and {Monte}, C. and {Monzani}, M.~E. and {Morselli}, A. and {Moskalenko}, I.~V. and {Mueller}, M. and {Murgia}, S. and {Nolan}, P.~L. and {Norris}, J.~P. and {Nuss}, E. and {Ohno}, M. and {Ohsugi}, T. and {Omodei}, N. and {Orlando}, E. and {Ormes}, J.~F. and {Ozaki}, M. and {Panetta}, J.~H. and {Parent}, D. and {Pelassa}, V. and {Pepe}, M. and {Pesce-Rollins}, M. and {Piron}, F. and {Porter}, T.~A. and {Rain{\`o}}, S. and {Rando}, R. and {Razzano}, M. and {Reimer}, A. and {Reimer}, O. and {Ritz}, S. and {Rodriguez}, A.~Y. and {Romani}, R.~W. and {Roth}, M. and {Ryde}, F. and {Sadrozinski}, H.~F.-W. and {Sander}, A. and {Scargle}, J.~D. and {Sgr{\`o}}, C. and {Shaw}, M.~S. and {Smith}, P.~D. and {Spandre}, G. and {Spinelli}, P. and {Starck}, J.-L. and {Strickman}, M.~S. and {Suson}, D.~J. and {Takahashi}, H. and {Takahashi}, T. and {Tanaka}, T. and {Thayer}, J.~B. and {Thayer}, J.~G. and {Thompson}, D.~J. and {Tibaldo}, L. and {Torres}, D.~F. and {Tosti}, G. and {Tramacere}, A. and {Uchiyama}, Y. and {Usher}, T.~L. and {Vasileiou}, V. and {Vilchez}, N. and {Vitale}, V. and {Waite}, A.~P. and {Wallace}, E. and {Wang}, P. and {Winer}, B.~L. and {Wood}, K.~S. and {Yang}, Z. and {Ylinen}, T. and {Ziegler}, M.},
        title = "{Gamma-ray Light Curves and Variability of Bright Fermi-detected Blazars}",
      journal = {\apj},
         year = 2010,
        month = oct,
       volume = {722},
       number = {1},
        pages = {520-542},
          doi = {10.1088/0004-637X/722/1/520},
archivePrefix = {arXiv},
       eprint = {1004.0348},
 primaryClass = {astro-ph.HE},
       adsurl = {https://ui.adsabs.harvard.edu/abs/2010ApJ...722..520A}
}

@ARTICLE{shah2025,
       author = {{Shah}, Zahir and {Dar}, Athar A. and {Akbar}, Sikandar and {Peer}, Anjum and {Malik}, Zahoor and {Manzoor}, Aaqib and {Ahanger}, Sajad and {Tantry}, Javaid and {Nazir}, Zeeshan and {Bose}, Debanjan and {Magray}, Mushtaq},
        title = "{Comprehensive variability analysis of blazars using Fermi light curves across multiple timescales}",
      journal = {\prd},
         year = 2025,
        month = jun,
       volume = {111},
       number = {12},
          eid = {123052},
        pages = {123052},
          doi = {10.1103/61tz-jk8c},
archivePrefix = {arXiv},
       eprint = {2505.23645},
 primaryClass = {astro-ph.HE},
       adsurl = {https://ui.adsabs.harvard.edu/abs/2025PhRvD.111l3052S}
}

@ARTICLE{press1978,
       author = {{Press}, W.~H.},
        title = "{Flicker noises in astronomy and elsewhere.}",
      journal = {Comments on Astrophysics},
         year = 1978,
        month = jan,
       volume = {7},
       number = {4},
        pages = {103-119},
       adsurl = {https://ui.adsabs.harvard.edu/abs/1978ComAp...7..103P}
}

@article{anderson15,
author = {Anderson, David F. and Ermentrout, Bard and Thomas, Peter J.},
title = {Stochastic representations of ion channel kinetics and exact stochastic simulation of neuronal dynamics},
year = {2015},
issue_date = {February  2015},
publisher = {Springer-Verlag},
address = {Berlin, Heidelberg},
volume = {38},
number = {1},
issn = {0929-5313},
url = {https://doi.org/10.1007/s10827-014-0528-2},
doi = {10.1007/s10827-014-0528-2},
journal = {J. Comput. Neurosci.},
month = feb,
pages = {67–82},
numpages = {16}
}

@article{gillespie76,
title = {A general method for numerically simulating the stochastic time evolution of coupled chemical reactions},
journal = {Journal of Computational Physics},
volume = {22},
number = {4},
pages = {403-434},
year = {1976},
issn = {0021-9991},
doi = {https://doi.org/10.1016/0021-9991(76)90041-3},
url = {https://www.sciencedirect.com/science/article/pii/0021999176900413},
author = {Daniel T Gillespie}
}

@article{gillespie77,
author = {Gillespie, Daniel T.},
title = {Exact stochastic simulation of coupled chemical reactions},
journal = {The Journal of Physical Chemistry},
volume = {81},
number = {25},
pages = {2340-2361},
year = {1977},
doi = {10.1021/j100540a008},
URL = {https://doi.org/10.1021/j100540a008
},
eprint = {https://doi.org/10.1021/j100540a008
}
}

@article{giannios13,
       author = {{Giannios}, Dimitrios},
        title = "{Reconnection-driven plasmoids in blazars: fast flares on a slow envelope}",
      journal = {\mnras},
         year = 2013,
        month = may,
       volume = {431},
       number = {1},
        pages = {355-363},
          doi = {10.1093/mnras/stt167},
archivePrefix = {arXiv},
       eprint = {1211.0296},
 primaryClass = {astro-ph.HE},
       adsurl = {https://ui.adsabs.harvard.edu/abs/2013MNRAS.431..355G}
}

@ARTICLE{petropolou16,
       author = {{Petropoulou}, Maria and {Giannios}, Dimitrios and {Sironi}, Lorenzo},
        title = "{Blazar flares powered by plasmoids in relativistic reconnection}",
      journal = {\mnras},
         year = 2016,
        month = nov,
       volume = {462},
       number = {3},
        pages = {3325-3343},
          doi = {10.1093/mnras/stw1832},
archivePrefix = {arXiv},
       eprint = {1606.07447},
 primaryClass = {astro-ph.HE},
       adsurl = {https://ui.adsabs.harvard.edu/abs/2016MNRAS.462.3325P}
}

@article{voytek15,
	author = {Voytek, Bradley and Kramer, Mark A. and Case, John and Lepage, Kyle Q. and Tempesta, Zechari R. and Knight, Robert T. and Gazzaley, Adam},
	title = {Age-Related Changes in 1/f Neural Electrophysiological Noise},
	volume = {35},
	number = {38},
	pages = {13257--13265},
	year = {2015},
	doi = {10.1523/JNEUROSCI.2332-14.2015},
	publisher = {Society for Neuroscience},
	issn = {0270-6474},
	URL = {https://www.jneurosci.org/content/35/38/13257},
	eprint = {https://www.jneurosci.org/content/35/38/13257.full.pdf},
	journal = {Journal of Neuroscience}
}

@article{btw87,
  title = {Self-organized criticality: An explanation of the 1/f noise},
  author = {Bak, Per and Tang, Chao and Wiesenfeld, Kurt},
  journal = {Phys. Rev. Lett.},
  volume = {59},
  issue = {4},
  pages = {381--384},
  numpages = {0},
  year = {1987},
  month = {Jul},
  publisher = {American Physical Society},
  doi = {10.1103/PhysRevLett.59.381},
  url = {https://link.aps.org/doi/10.1103/PhysRevLett.59.381}
}

@article{halley96,
title = {Ecology, evolution and 1f-noise},
journal = {Trends in Ecology \& Evolution},
volume = {11},
number = {1},
pages = {33-37},
year = {1996},
issn = {0169-5347},
doi = {https://doi.org/10.1016/0169-5347(96)81067-6},
url = {https://www.sciencedirect.com/science/article/pii/0169534796810676},
author = {John M. Halley}
}

@article{schottky25,
  title = {The Schottky Effect in Low Frequency Circuits},
  author = {Johnson, J. B.},
  journal = {Phys. Rev.},
  volume = {26},
  issue = {1},
  pages = {71--85},
  numpages = {0},
  year = {1925},
  month = {Jul},
  publisher = {American Physical Society},
  doi = {10.1103/PhysRev.26.71},
  url = {https://link.aps.org/doi/10.1103/PhysRev.26.71}
}

@ARTICLE{smol16,
       author = {{Smoluchowski}, M.~V.},
        title = "{Drei Vortrage uber Diffusion, Brownsche Bewegung und Koagulation von Kolloidteilchen}",
      journal = {Zeitschrift fur Physik},
         year = 1916,
        month = jan,
       volume = {17},
        pages = {557-585},
       adsurl = {https://ui.adsabs.harvard.edu/abs/1916ZPhy...17..557S}
}

@article{lewis79,
author = {Lewis, P. A. W and Shedler, G. S.},
title = {Simulation of nonhomogeneous poisson processes by thinning},
journal = {Naval Research Logistics Quarterly},
volume = {26},
number = {3},
pages = {403-413},
doi = {https://doi.org/10.1002/nav.3800260304},
url = {https://onlinelibrary.wiley.com/doi/abs/10.1002/nav.3800260304},
eprint = {https://onlinelibrary.wiley.com/doi/pdf/10.1002/nav.3800260304},
year = {1979}
}

@ARTICLE{fermo2010,
       author = {{Fermo}, R.~L. and {Drake}, J.~F. and {Swisdak}, M.},
        title = "{A statistical model of magnetic islands in a current layer}",
      journal = {Physics of Plasmas},
         year = 2010,
        month = jan,
       volume = {17},
       number = {1},
          eid = {010702},
        pages = {010702},
          doi = {10.1063/1.3286437},
archivePrefix = {arXiv},
       eprint = {0910.4971},
 primaryClass = {physics.plasm-ph},
       adsurl = {https://ui.adsabs.harvard.edu/abs/2010PhPl...17a0702F}
}

@ARTICLE{marsch85,
       author = {{Marscher}, A.~P. and {Gear}, W.~K.},
        title = "{Models for high-frequency radio outbursts in extragalactic sources, with application to the early 1983 millimeter-to-infrared flare of 3C 273.}",
      journal = {\apj},
         year = 1985,
        month = nov,
       volume = {298},
        pages = {114-127},
          doi = {10.1086/163592},
       adsurl = {https://ui.adsabs.harvard.edu/abs/1985ApJ...298..114M}
}

@ARTICLE{gian09,
       author = {{Giannios}, Dimitrios and {Uzdensky}, Dmitri A. and {Begelman}, Mitchell C.},
        title = "{Fast TeV variability in blazars: jets in a jet}",
      journal = {\mnras},
         year = 2009,
        month = may,
       volume = {395},
       number = {1},
        pages = {L29-L33},
          doi = {10.1111/j.1745-3933.2009.00635.x},
archivePrefix = {arXiv},
       eprint = {0901.1877},
 primaryClass = {astro-ph.HE},
       adsurl = {https://ui.adsabs.harvard.edu/abs/2009MNRAS.395L..29G}
}

@ARTICLE{marscher14,
       author = {{Marscher}, Alan P.},
        title = "{Turbulent, Extreme Multi-zone Model for Simulating Flux and Polarization Variability in Blazars}",
      journal = {\apj},
         year = 2014,
        month = jan,
       volume = {780},
       number = {1},
          eid = {87},
        pages = {87},
          doi = {10.1088/0004-637X/780/1/87},
archivePrefix = {arXiv},
       eprint = {1311.7665},
 primaryClass = {astro-ph.HE},
       adsurl = {https://ui.adsabs.harvard.edu/abs/2014ApJ...780...87M}
}

@ARTICLE{raiteri25,
       author = {{Raiteri}, Claudia M.},
        title = "{The variability of blazars throughout the electromagnetic spectrum}",
      journal = {arXiv e-prints},
         year = 2025,
        month = nov,
          eid = {arXiv:2511.18975},
        pages = {arXiv:2511.18975},
          doi = {10.48550/arXiv.2511.18975},
archivePrefix = {arXiv},
       eprint = {2511.18975},
 primaryClass = {astro-ph.HE},
       adsurl = {https://ui.adsabs.harvard.edu/abs/2025arXiv251118975R}
}

@ARTICLE{sinha18,
       author = {{Sinha}, Atreyee and {Khatoon}, Rukaiya and {Misra}, Ranjeev and {Sahayanathan}, Sunder and {Mandal}, Soma and {Gogoi}, Rupjyoti and {Bhatt}, Nilay},
        title = "{The flux distribution of individual blazars as a key to understand the dynamics of particle acceleration}",
      journal = {\mnras},
         year = 2018,
        month = oct,
       volume = {480},
       number = {1},
        pages = {L116-L120},
          doi = {10.1093/mnrasl/sly136},
archivePrefix = {arXiv},
       eprint = {1807.09073},
 primaryClass = {astro-ph.HE},
       adsurl = {https://ui.adsabs.harvard.edu/abs/2018MNRAS.480L.116S}
}

@ARTICLE{biteau12,
       author = {{Biteau}, J. and {Giebels}, B.},
        title = "{The minijets-in-a-jet statistical model and the rms-flux correlation}",
      journal = {\aap},
         year = 2012,
        month = dec,
       volume = {548},
          eid = {A123},
        pages = {A123},
          doi = {10.1051/0004-6361/201220056},
archivePrefix = {arXiv},
       eprint = {1210.2045},
 primaryClass = {astro-ph.HE},
       adsurl = {https://ui.adsabs.harvard.edu/abs/2012A&A...548A.123B}
}

@article{mw47,
  author = {Mann, Henry B. and Whitney, Donald R.},
  title = {On a Test of Whether one of Two Random Variables is Stochastically Larger than the Other},
  journal = {Annals of Mathematical Statistics},
  volume = {18},
  number = {1},
  pages = {50--60},
  year = {1947},
  publisher = {Institute of Mathematical Statistics}
}
\bibliographystyle{aasjournalv7}



\end{document}